\documentclass{emulateapj}
\usepackage{apjfonts}

\usepackage{graphicx}
\usepackage{amsmath}
\usepackage{amssymb}
\usepackage[]{natbib}
\usepackage[usenames]{color}
\usepackage{float}
\usepackage{subfigure}
\usepackage{hyperref}

\begin{document}

\title{Forming Double-barred Galaxies From Dynamically Cool Inner Disks}
\author{
Min Du\altaffilmark{1,2},
Juntai Shen\altaffilmark{1,3},
Victor P. Debattista\altaffilmark{4}}

\altaffiltext{1}{Key Laboratory of Research in Galaxies and Cosmology, Shanghai Astronomical Observatory, Chinese Academy of Sciences, 80 Nandan Road, Shanghai 200030, China}
\altaffiltext{2}{University of China Academy of Sciences, 19A Yuquanlu, Beijing 100049, China}
\altaffiltext{3}{Correspondence should be addressed to Juntai Shen: jshen@shao.ac.cn}

\altaffiltext{4}{Jeremiah Horrocks Institute, University of Central Lancashire, Preston, PR1 2HE, UK}

\begin{abstract}

About one third of early-type barred galaxies host small-scale
secondary bars. The formation and evolution of such double-barred
galaxies remain far from being well understood. In order to understand 
the formation of such systems, we explore a large parameter space of 
isolated pure-disk simulations. We show that a 
dynamically cool inner disk embedded in a hotter outer disk can naturally
generate a steady secondary bar while the outer disk forms a large-scale 
primary bar. The independent bar instabilities of inner and outer disks 
result in long-lived double-barred structures whose dynamical properties 
are comparable with observations. This formation scenario indicates that 
the secondary bar might form from the general bar instability, the same as
the primary bar. Under some circumstances, the interaction of the bars 
and the disk leads to the two bars aligning or single, nuclear, bars only. 
Simulations that are cool enough of the center to experience clump 
instabilities may also generate steady double-barred 
galaxies. In this case, the secondary bars are ``fast'', i.e., the bar 
length is close to the co-rotation radius. This is the first time that 
double-barred galaxies containing a fast secondary bar are reported. 
Previous orbit-based studies had suggested that fast secondary bars 
are not dynamically possible.

\end{abstract}



\section{Introduction}
Double-barred (S2B) galaxies, consisting of a small-scale secondary bar
embedded in its large-scale primary counterpart, were first described nearly 40
years ago \citep{de_75}. Statistics of the fraction of S2Bs amongst
early-type galaxies have been obtained from optical \citep{erw_spa_02, erw_04} 
and infrared \citep{lai_etal_02} observations which showed that about one 
third of early-type barred galaxies are S2Bs. We still lack systematic surveys 
of later Hubble types because of their stronger dust extinction, especially 
in the central region \citep{erw_05}.

Observations \citep{but_cro_93,fri_mar_93, cor_etal_03} have 
shown that the two bars rotate independently, which was expected from 
numerical simulations \citep[e.g.][hereafter DS07]{shl_hel_02,deb_she_07}.
In general, nested bars cannot rotate through each other rigidly
\citep{lou_ger_88}. In the potential of two independently rotating bars, 
the orbits may not be closed in any reference frame. 
\citet{mac_spa_97,mac_spa_00} and \citet{mac_ath_07} studied the orbits based 
on the concept of {\it loops}, which is a family of orbits whose population of
particles return to the same curve, but not to the same position, when the
two bars return to the same relative orientation. They also showed
non-rigid rotation for {\it loops}. Dynamically decoupled secondary bars
in S2Bs have been hypothesized to be a mechanism for driving gas past
the inner Lindblad resonance (ILR) of the primary bars to feed the
supermassive black holes that power active galactic nuclei (AGN)
\citep{shl_etal_89,shl_etal_90}.

Other works studied S2Bs from a purely kinematical point of view.
\citet{de_etal_08} presented the 2-D stellar velocity and velocity
dispersion maps of a sample of four S2Bs, based on observations
with the \texttt{SAURON} integral-field spectrograph.
The high quality velocity dispersion maps reveal two local minima, located
near the ends of the secondary bar of each galaxy. They suggested that
these $\sigma$-hollows appear because of the contrast between
the velocity dispersion of a hotter bulge and the secondary bar, as the
secondary bar is dominated by ordered motion and thus has a low~$\sigma$.

The formation of S2Bs has been studied by numerical simulations. The 
best-known scenario for forming independently rotating double bars was
proposed by \citet{fri_mar_93}: a pre-existing large-scale bar
drives gas inflow into the central kiloparsec of a galaxy; Once
sufficient gas has accumulated, it becomes bar-unstable and a dynamically
decoupled (gaseous) secondary bar forms \citep[see also][]{com_94, shl_hel_02, eng_shl_04}. 
However, the S2B structures forming from gas are short-lived and gas 
dominated; this cannot explain the observed high abundance of S2Bs in 
gas-poor early-type galaxies \citep{pet_wil_04}. However, recently 3-D 
N-body+hydrodynamical simulations by \citet{woz_15} formed long-lived S2Bs 
that might provide a better description of this process. The new stars 
form from central gas accumulations producing a dynamically cool inner disk. 
The stellar populations are in qualitative agreement with 
the observations of \citet{de_etal_13} in that the secondary bars are a few 
Gyr younger than their primary counterparts.

\citet{rau_sal_99} and \citet{rau_etal_02} reported the formation of S2Bs 
in purely collisionless studies, although these secondary bars often have 
a ``vaguely spiral shape''. DS07 and \citet{she_deb_09} performed 
well-resolved simulations of long-lived S2Bs with a pre-existing rapidly 
rotating pseudobulge without any gas. Finally, some simulations have 
indicated that the dark matter halo may sometimes play a role in generating 
S2Bs \citep{hel_etal_07, sah_mac_13}. Thus the conditions of S2B formation 
are still not well understood. Nevertheless, this variety of simulations 
shows that gas is not required to form secondary bars. 

Here we present new three-dimensional (3D) N-body simulations that 
successfully generate S2Bs from simple and natural initial conditions. We form
long-lived S2B structures in pure-disk simulations by choosing different 
initial parameters for the inner and outer parts of the disk. Starting from
different dynamical conditions, the inner and outer disks generate independent 
bar instabilities that result in a high probability of S2B formation. The 
paper is organised as follows. In Section \ref{sectionsetup} we describe the
model setup. In Section \ref{sectionoutcome} we summarize the results of 
exploring the parameter space. 
In Section \ref{sectionS2B} we show case studies of double-barred 
galaxies. A comparison with single-barred galaxies is presented in Section 
\ref{sectionsingle}. Finally, in Section \ref{sectionconclusion} we 
summarize and discuss the implications of this work.

\section{Model setup}
\label{sectionsetup}
The N-body simulations are evolved with a 3D cylindrical polar grid code, 
\texttt{GALAXY} \citep{sel_val_97, sel_14}. All models consist of a live 
exponential disk (scale length $R_d$, stellar mass $M_d$) and a rigid 
dark matter halo, with no 
gas present. The simulations use units where $G=M_0=R_d=V_0=T_0=1$. The 
key point is that a dynamically cool inner disk is used to generate 
strong small-scale bar instabilities. As shown in Eq. \ref{eq.Q}, the initial 
Toomre-$Q$ profile is roughly constant at $2.0$ in the outer disk ($R>1.75$), 
while in the inner disk it decreases gradually to $b_Q$ following a quadratic 
curve:
\begin{equation}\label{eq.Q}
      Q(R)=  
      \left\{
      \begin{aligned}
            & (2.0-b_Q)(\frac{R}{1.75})^2 +b_Q & (R\leq1.75) \\
            & 2.0 & (R>1.75). 
      \end{aligned}
      \right.
\end{equation}
Thus the outer disk is dynamically hot, while the inner part of the disk is 
dominated by ordered motion. The only free parameter, $b_Q$, can be used to 
set the dynamical temperature of the inner disk. The initial thickness 
$z_0$ is $0.1$ all over the disk. 

In this study, the halo potential is logarithmic 
\begin{equation}
        \Phi(r)=\frac{1}{2}{V_h^2}\ln(r^2+r_h^2),
\end{equation}
with $V_h=0.6$ and $r_h=15$. 
Because the models are very disk dominated, the halo density is low near 
the center where the secondary bar dominates. We expect that the angular momentum 
transfer between the bar and the halo should be less important than models in 
\citet{deb_sel_00} and \citet{ath_03}. Then rigid halos are useful for studying 
the complicated co-evolution of the two bars without the additional weaker 
evolution introduced by a live halo, by allowing high mass resolution in 
the nuclear regions. The initial disk has $4\times10^6$ equal-mass particles. A 
possible scaling to physical values is $M_0=4.0\times10^{10}M_{\odot}$ 
and $R_d=3.0$ kpc, which gives a velocity unit of 
$V_0={(\frac{GM_0}{R_d})}^{1/2}\simeq 239$ km/s and a time unit of 
$T_0=R_d/V_0\simeq 12.3$ Myr. The force resolution 
(softening) is 0.01 (corresponding to $30$ pc). 
The forces in the radial direction are solved
by direct convolution with the Greens function, while the vertical 
and azimuthal forces are obtained by fast Fourier transform.
We use grids measuring $N_R \times N_{\phi} \times N_z = 58 \times 64
\times 375$ but have verified that increasing resolution does not affect
our results by doubling $N_R$ and $N_{\phi}$. The vertical spacing of 
the grid planes was $\delta z = 0.01$. Time integration used a leapfrog 
integrator with a fixed time step $\delta t=0.04$ corresponding to about 
$0.5$ Myr.

\section{Exploring the parameter space}
\label{sectionoutcome}
\begin{figure}[htp]
        \centering
        \subfigure{\includegraphics[width=0.48\textwidth]{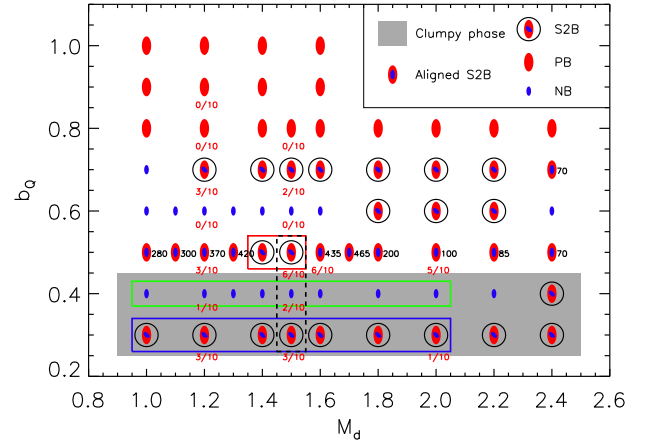}}
        \caption{Exploring the parameter space of $b_Q$ and $M_d$.
            The long-lived double-barred, single-barred and nuclear-barred 
            simulations are marked as ``S2B'', ``PB'' and ``NB'', 
            respectively. The simulations whose two bars align with each other 
            during the simulation are marked as ``aligned S2B''. 
            The black numbers at the right side of aligned S2Bs are the 
            times when the alignment finishes. The shaded region shows the 
            range of $b_Q$ and $M_d$ when the inner disk suffers strong clump 
            instabilities at the beginning. 
            The red, green, and blue rectangles highlight the subsamples of 
            standard S2B, clumpy S2B and NB models, while the dashed black
            rectangle marks the three models with the same disk mass to 
            compare their evolution of their surface density. The results of 
            the stochasticity test are shown as the fractions in red.}
        \label{fig:parspace}
\end{figure}
Based on linear bar-formation theory \citep{too_81}, the modes of
bar formation are standing waves in a cavity, akin to the familiar modes of
organ pipes and guitar strings. 
To form double-barred systems, we strive to build a disk with 
independent bar instabilities in the inner and outer regions. The 
responsiveness of the disk can be enhanced by decreasing $Q$,  
increasing surface density and reducing disk thickness \citep{sel_89}. 
Therefore, we explore the parameter space of $b_Q$, $M_d$ and thickness. 
A set of simulations (not presented here) show that reducing the thickness 
of the inner disk does not strongly affect the formation of S2Bs. 
The results of simulations with varying $b_Q$ and $M_d$ are shown in Figure 
\ref{fig:parspace}. We classify the outcomes into five types based on 
their formation history and final morphology. There are three types 
of S2Bs: standard S2Bs, clumpy S2Bs and aligned (or coupled) S2Bs. Aligned S2Bs 
are unstable S2Bs whose two bars eventually couple into alignment after a few 
Gyr. The steady S2Bs forming from a violent clumpy phase when 
$b_Q \lesssim 0.45$ are termed clumpy S2Bs. S2Bs not experiencing a clumpy 
phase are termed ``standard'' S2Bs. The two types of single-barred models 
are divided into large-scale primary single-barred models (PBs) and 
nuclear-barred models (NBs) depending on the size of the bar. If the 
semi-major axis is less than $1.0$, we classify it as a NB, otherwise, it is 
a PB. 

The chaotic nature of disks leads to significant stochasticity 
\citep{mil_64, sel_deb_09}. Because of the interaction of 
multiple non-axisymmetric components, the simulations presented here are 
more stochastic than typical simulations involving one bar. Therefore, we 
test the degree of stochasticity for a subsample of models by changing 
only the random seed when generating the particle initial conditions.
The results are shown as the fractions in red in Figure \ref{fig:parspace}, 
whose denominator corresponds to the total number of simulations we have run, 
while the numerator is the number of simulations forming 
steady S2Bs, including standard S2Bs and clumpy S2Bs. The fractions give 
an approximate probability for forming long-lived S2Bs. When S2Bs fail to 
form, the outcomes can be either aligned S2Bs, or single bars, either NBs 
or PBs. 

\section{The formation and evolution of double-barred galaxies}
\label{sectionS2B}

As shown in Figure \ref{fig:parspace}, the most important condition for 
S2B formation from independent bar instabilities is given by the $Q$ 
profile. When $b_Q\gtrsim0.8$, S2Bs do not form for a large range 
of disk mass. Instead most of such simulations form PBs only. 

\subsection{A standard double-barred galaxy}
\label{subsectionS2B1}
\begin{figure*}[htp]
        \centering
        \includegraphics[width=0.56\textwidth, angle=-90]{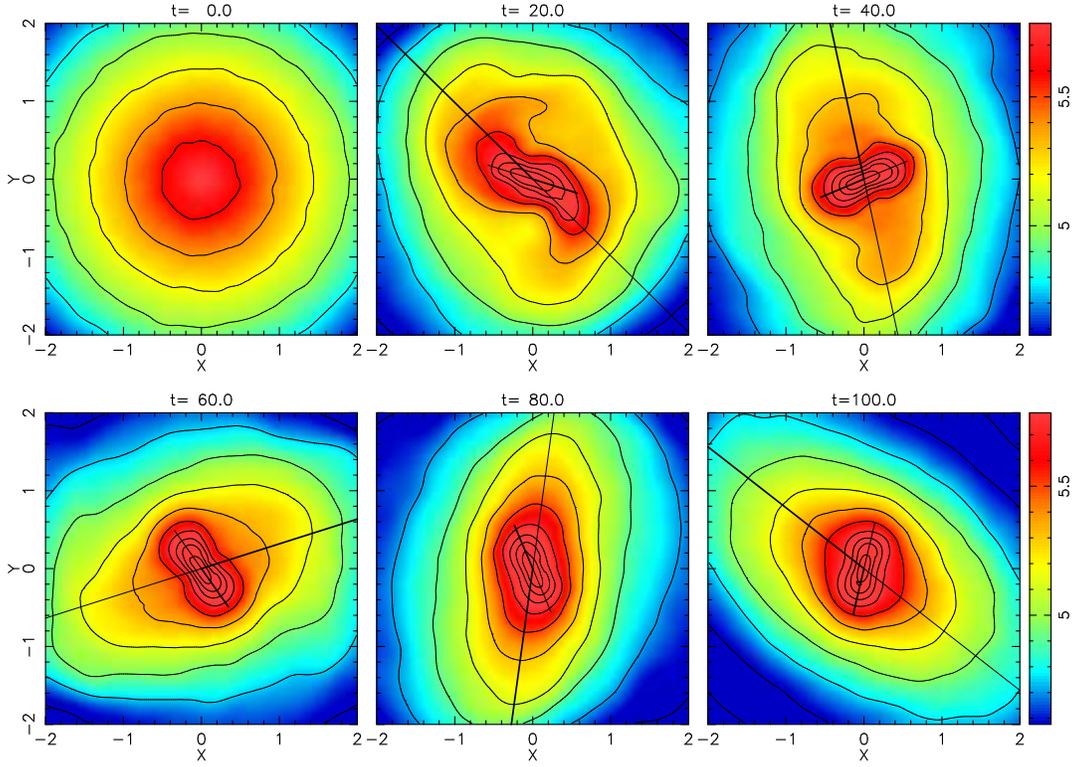}
        \caption{The face-on images of the S2B simulation at various times, 
            with isodensity contours superposed. The contours are 
            equally separated in logarithm, showing the formation process of 
            the S2B structure over the interval $0\le{t}\le 100$. The surface 
            density contours have been smoothed with an adaptive kernel \citep{sil_86}. 
            The short and long straight lines mark the major axes 
            of the secondary bar and the primary bar, respectively. A movie showing the evolution of the standard S2B in the primary bar's corotating 
            frame is available at \href{http://hubble.shao.ac.cn/~dumin/S2B/Video1.gif}{hubble.shao.ac.cn/$\sim$dumin/S2B/Video1.gif} and the ApJ website.}
        \label{fig:S2Bform}
\end{figure*}
\begin{figure*}[htp]
        \centering
        \subfigure{\includegraphics[width=0.20\textwidth, angle=-90]{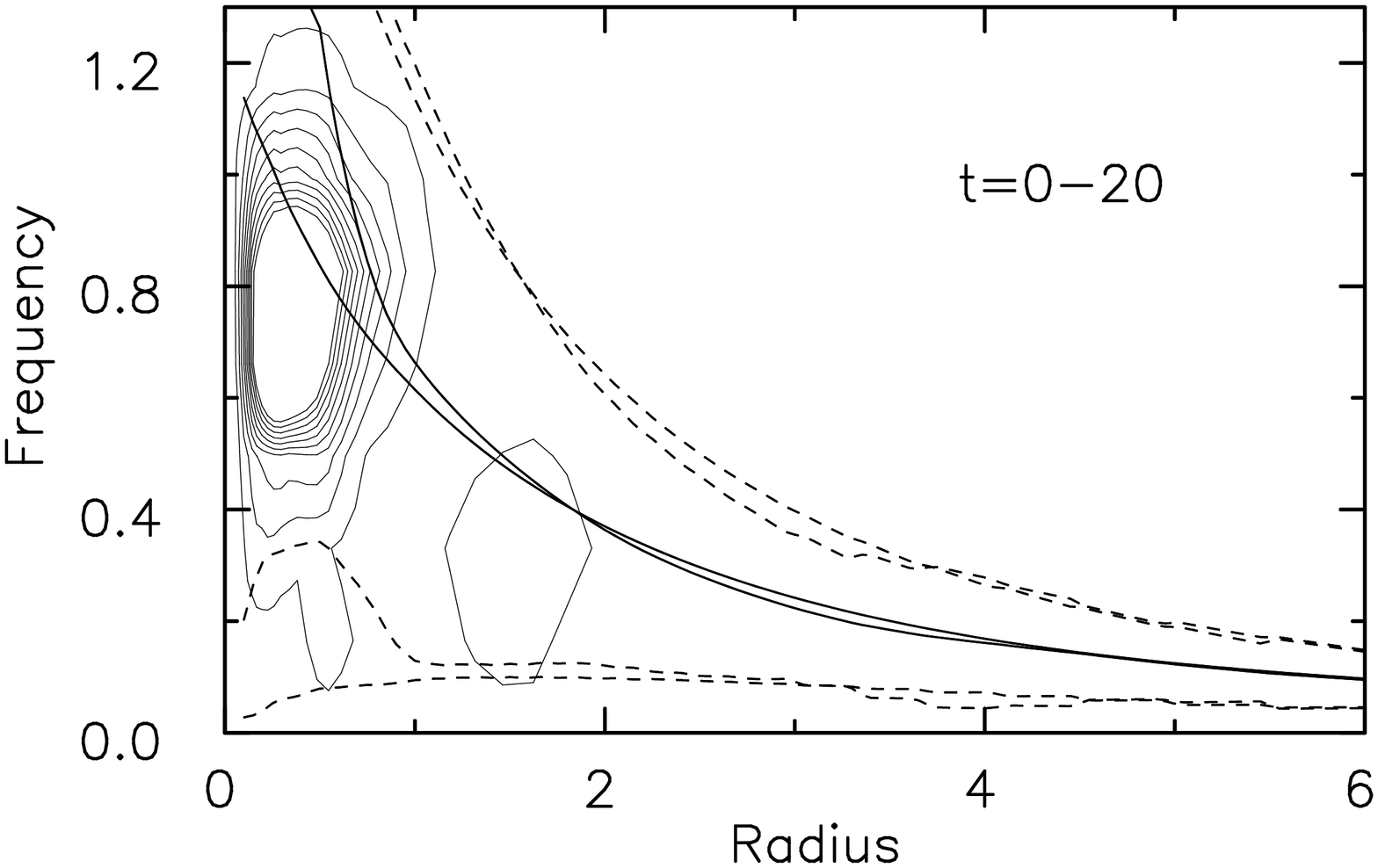}}
        \subfigure{\includegraphics[width=0.20\textwidth, angle=-90]{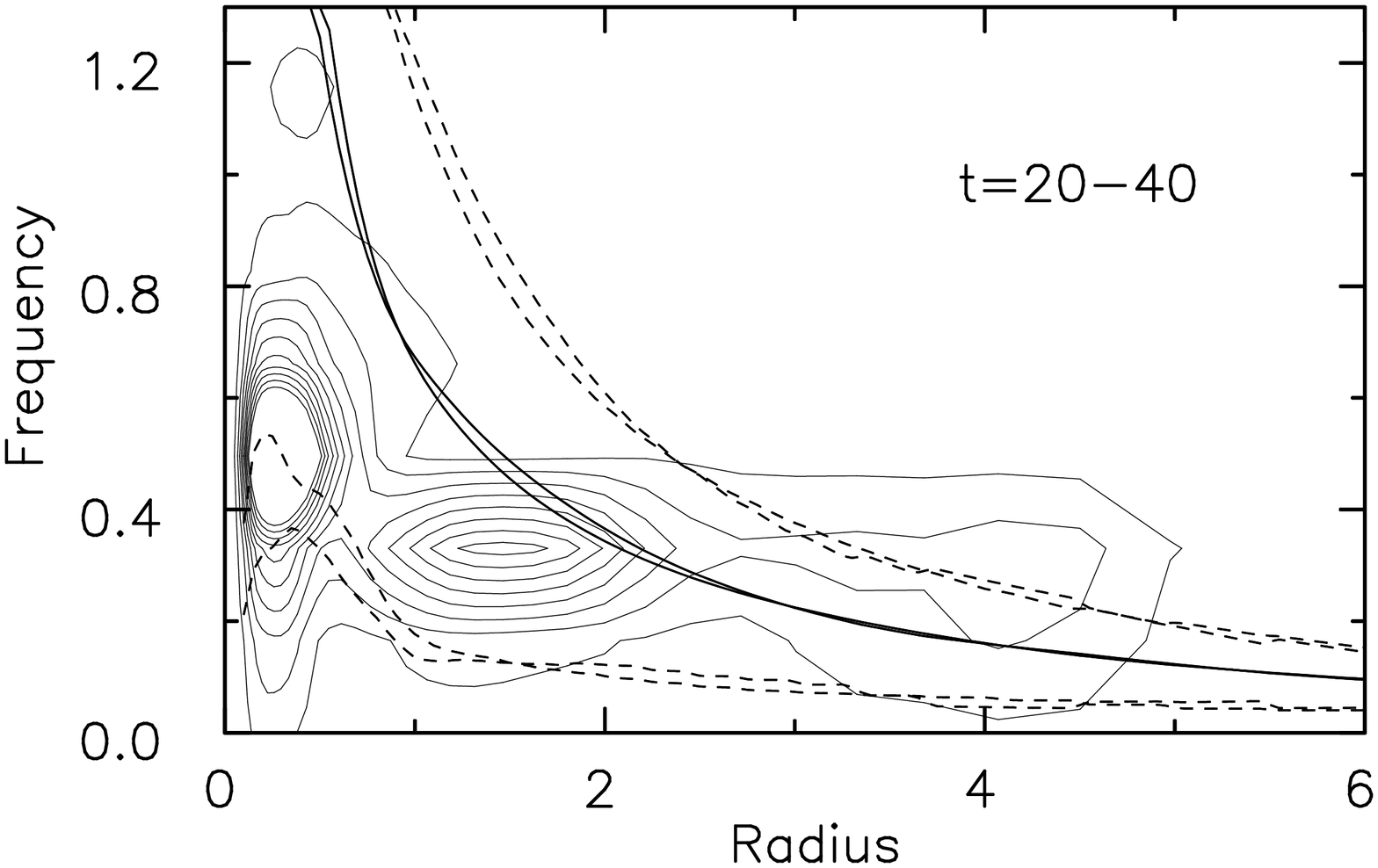}}
        \subfigure{\includegraphics[width=0.20\textwidth, angle=-90]{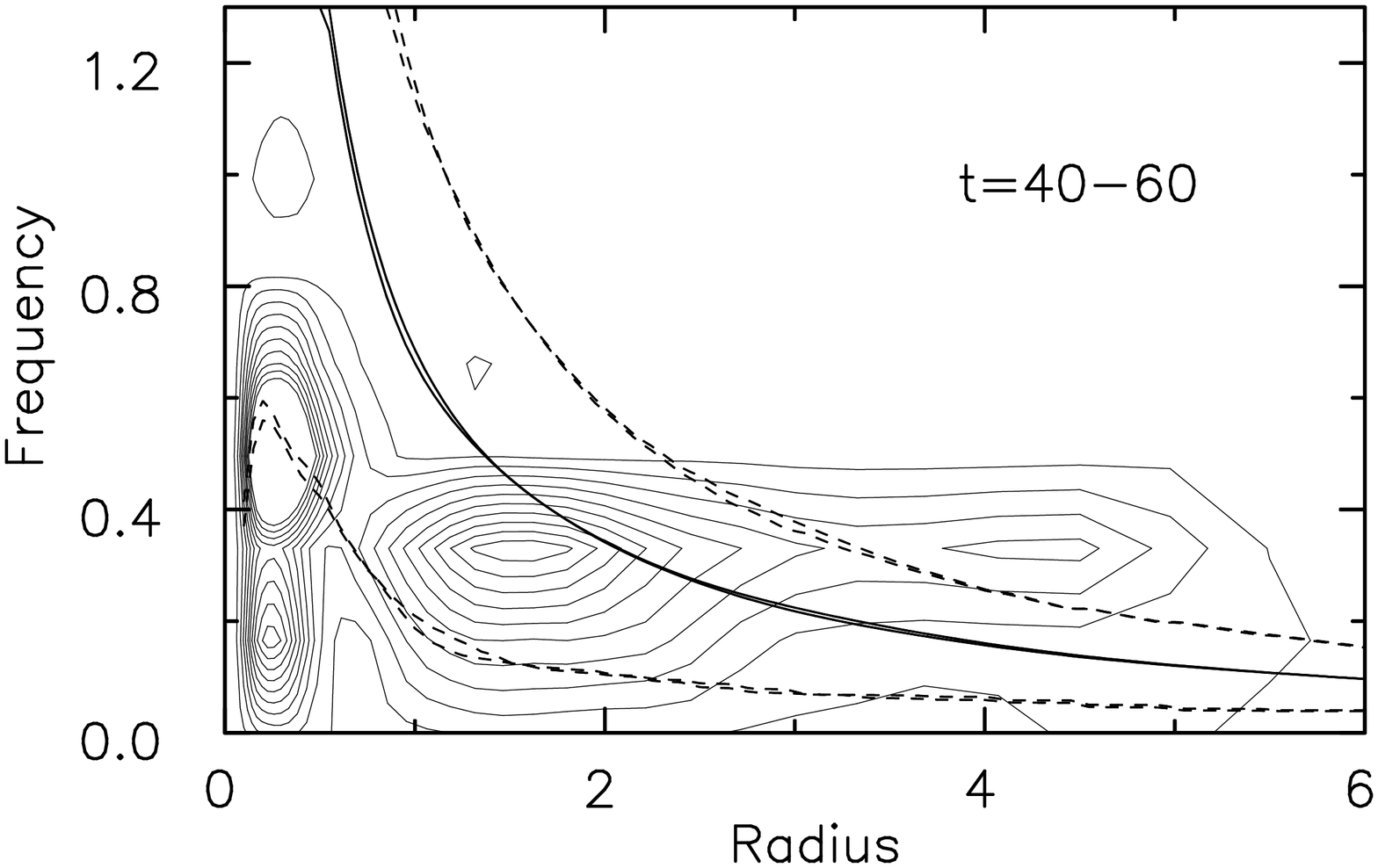}}
        \subfigure{\includegraphics[width=0.20\textwidth, angle=-90]{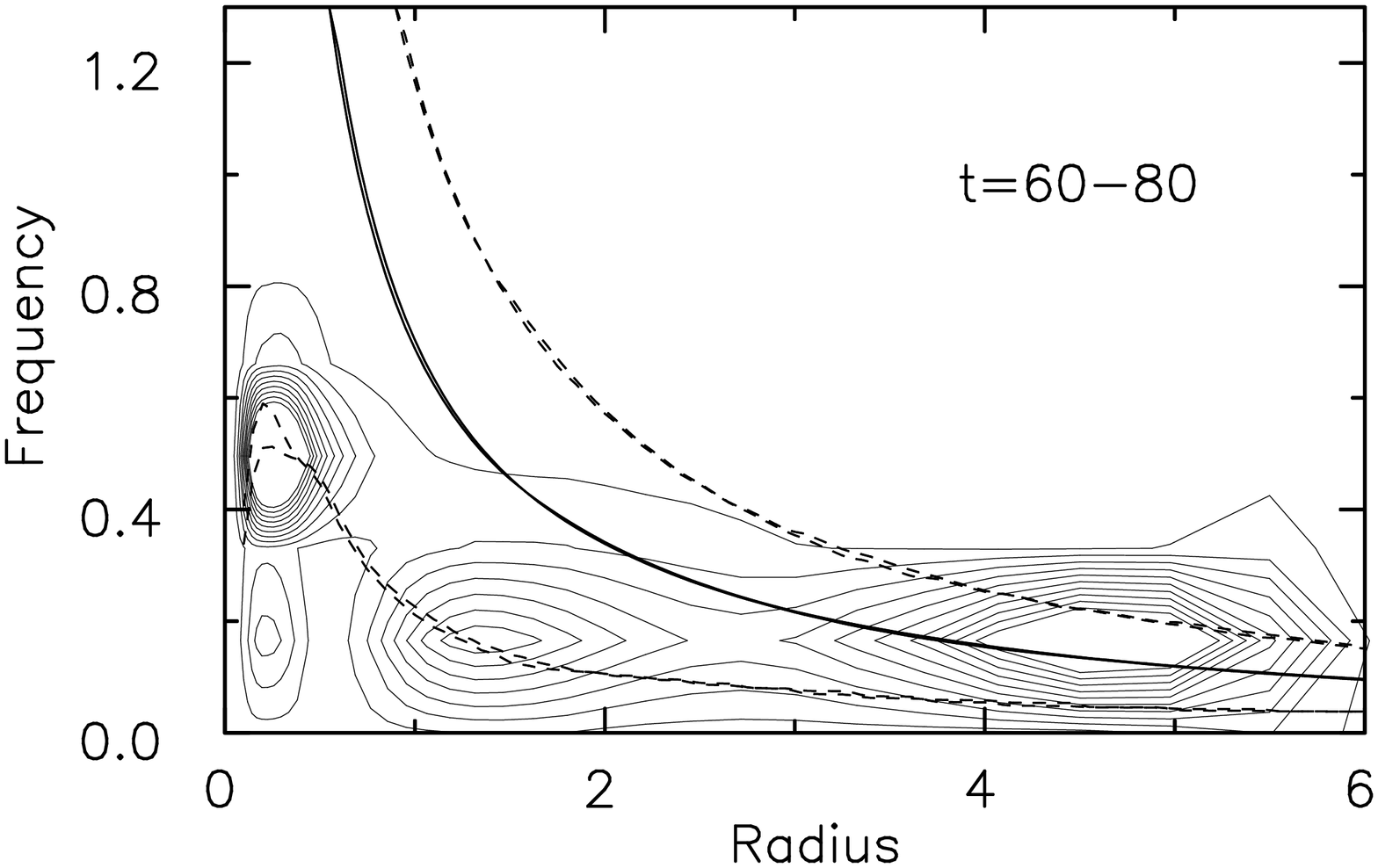}}
        \subfigure{\includegraphics[width=0.20\textwidth, angle=-90]{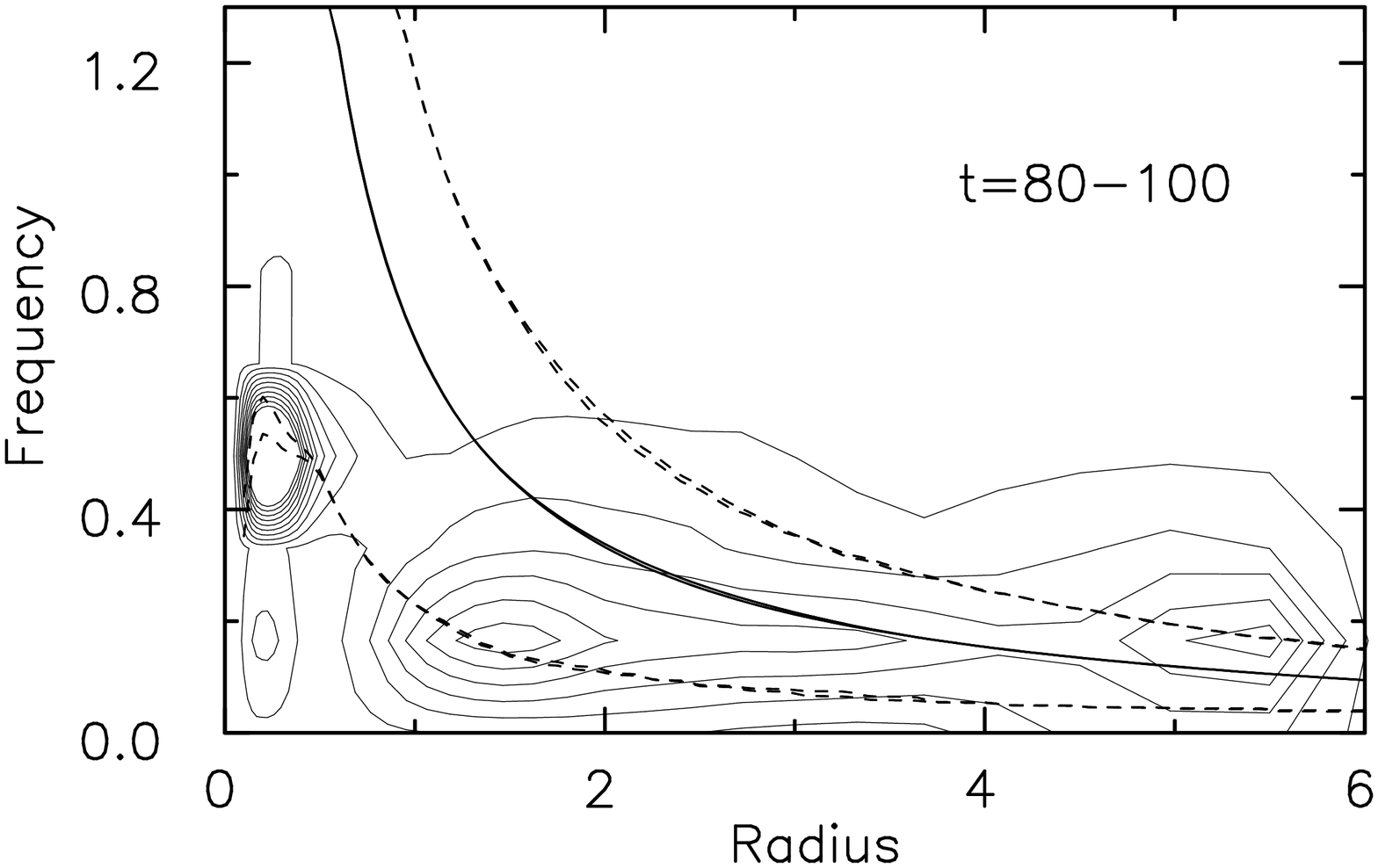}}
        \subfigure{\includegraphics[width=0.20\textwidth, angle=-90]{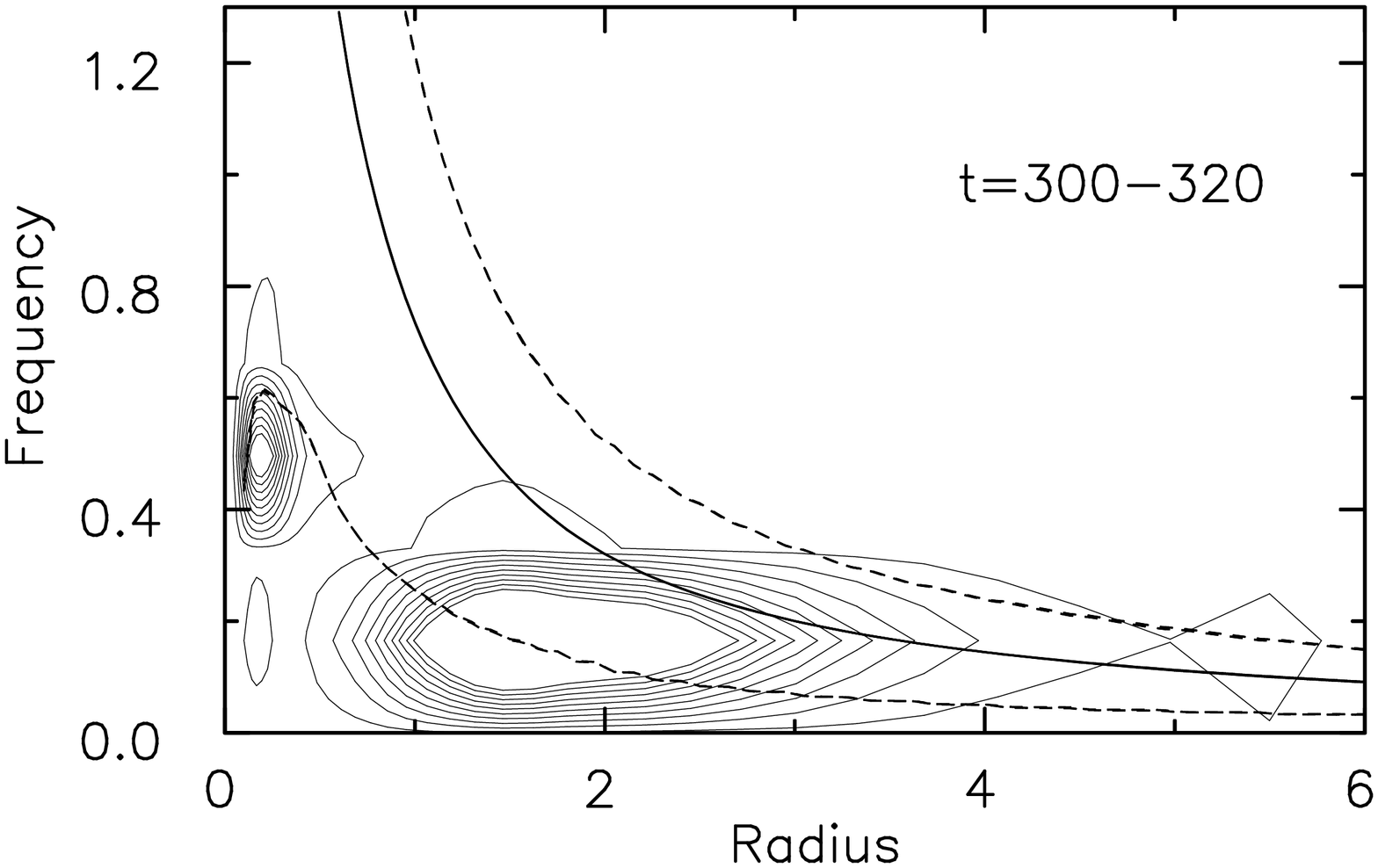}}
        \caption{Power spectra of the $m=2$ Fourier component as a 
        function of radius. 
        The signals of the $m=2$ sectoral harmonic of the density are 
        measured over a time interval of $20$. The solid lines show $\Omega(R)$, the 
        frequency of circular rotation. The dashed lines show 
        $\Omega + \kappa/2$ (upper) and $\Omega - \kappa/2$ (lower), 
        where $\kappa$ is the epicyclic frequency. The two sets of solid 
        and dashed lines correspond to the curves of the starting and ending 
        times in each interval, respectively. 
        Generally, the lower sets denote the starting times, while 
        the upper sets denote the ending times.}
        \label{fig:S2Bfreq}
\end{figure*}
\begin{figure}[htp]
        \centering
        \includegraphics[width=0.45\textwidth]{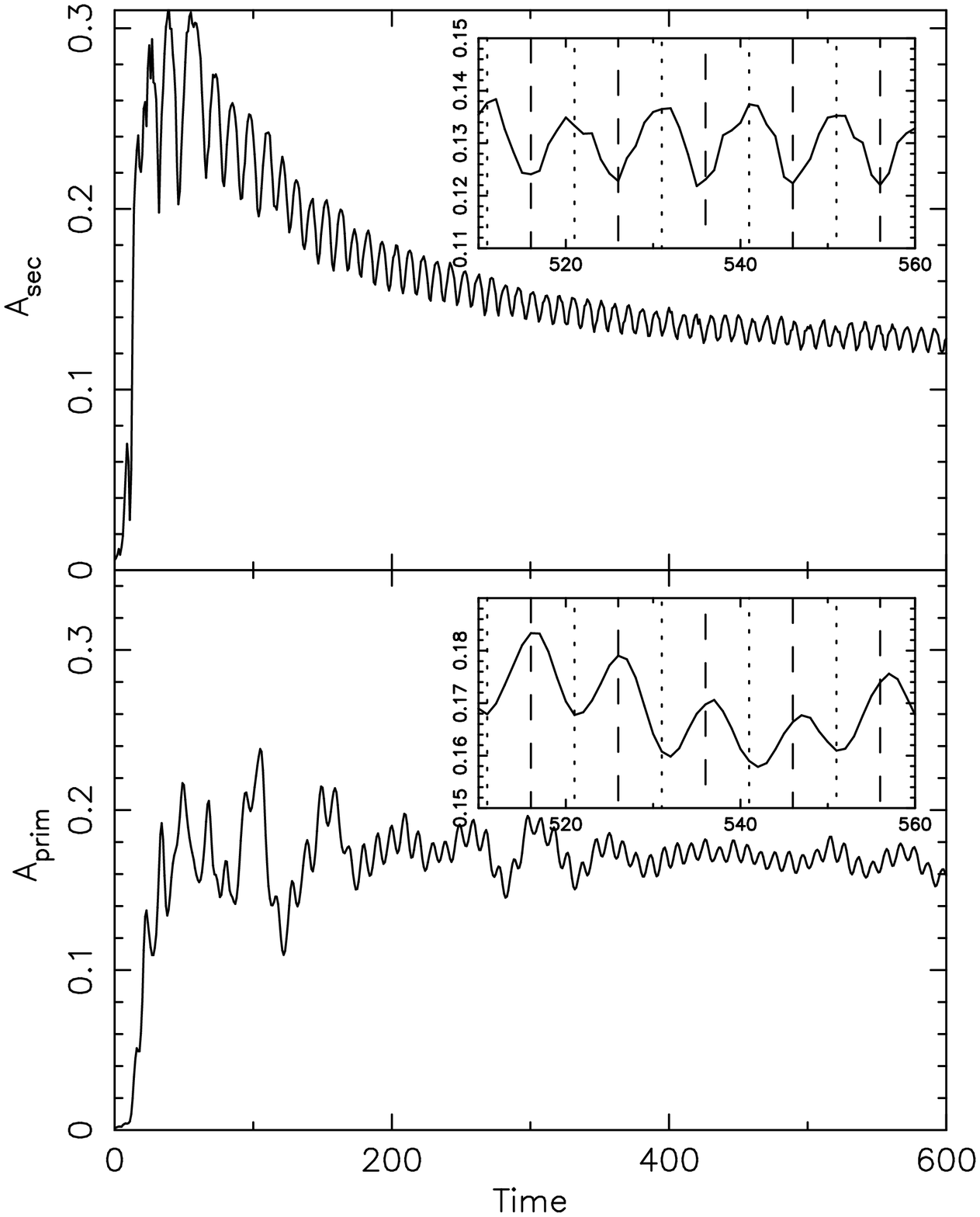}
        \caption{Time evolution of the amplitudes of the secondary bar 
        (top panel) and the primary bar (bottom panel) of the standard S2B. 
        $A_{\rm prim}$ and $A_{\rm sec}$ are defined as the Fourier $m=2$ 
        amplitude over the radial ranges $0.96\le{R}\le3.01 $ and 
        $R\leq0.30$, respectively. In the insets, the dashed lines mark 
        the times when the two bars are aligned, while the dotted lines 
        mark the times when they are perpendicular.}
        \label{fig:S2Bevo}
\end{figure}

The model shown in Figure \ref{fig:S2Bform} is a standard S2B. The initial 
central dynamical temperature parameter, $b_Q$, is set to $0.5$. In order 
to be sufficiently massive for bar instabilities, the mass of the disk is 
$M_d=1.5M_0=6.0\times10^{10}M_{\odot}$. Figures \ref{fig:S2Bform} and 
\ref{fig:S2Bfreq} give an overview of the formation of the S2B over 
$100$ time units ($1.23$ Gyr). The time evolution of the amplitudes of the 
primary bar ($A_{\rm prim}$) and the secondary bar ($A_{\rm sec}$) is shown 
in Figure \ref{fig:S2Bevo}. 

From both analytical calculations and N-body simulations, we know that 
the length, strength and pattern speed reduction of bars are strongly 
affected by the angular momentum exchange with the outer disk 
\citep[e.g.][]{lyn_kal_72, ath_03}. Because of the low initial $Q$, the 
inner disk starts with strong $m=2$ instabilities ($t=0-20$, Figure 
\ref{fig:S2Bfreq}) with high pattern speed $\Omega_{\rm sec} \sim 0.80$. 
A small bar forms extending roughly to its co-rotation radius $R_{CR} \sim 1.0$, 
which indicates that it forms via the usual bar instability \citep{too_81}. 
As time goes on, the small bar sheds angular momentum and traps particles 
from the disk further out, which happens rapidly as the dynamical time 
scale is short in the inner disk. In this period ($t=0-50$), $A_{\rm sec}$ 
constantly increases, while the pattern speed of the secondary bar declines to 
$\Omega_{\rm sec} \sim 0.50$ ($t=40-60$). 

\begin{figure}[htp]
    \centering
    \includegraphics[width=0.35\textwidth, angle=-90]{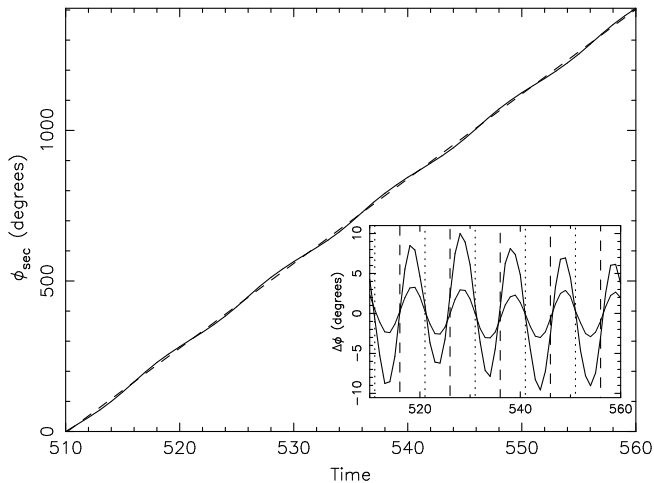}
    \caption{Time evolution of the phase of the secondary bar, measured 
    between $t=510-560$. The dashed straight line is the least-square 
    fit, which gives the slope $\langle \Omega_{\rm sec} \rangle$. The 
    inset figure shows the phase 
    difference, $\Delta \phi$, between the phases of the bars and 
    $\langle \Omega \rangle t$, where $\langle \Omega \rangle$ is the 
    pattern speed averaged over one relative rotation of the two bars; 
    the thick line with a large-amplitude oscillation is for the secondary 
    bar while the thin line with a small-amplitude oscillation is for the 
    primary bar. The dashed (dotted) lines mark the times that the two 
    bars are parallel (perpendicular) to each other as 
    in Figure \ref{fig:S2Bevo}.}
    \label{fig:S2Bphase}
\end{figure}

The primary bar forms a little later and slower than the secondary bar 
through a quite similar process. It starts with $\Omega_{\rm prim} \sim 0.37$ 
($t=20-40$) and declines to $\sim 0.18$ at $t=60-80$. 
From the evolution of 
$A_{\rm sec}$ (Figure \ref{fig:S2Bevo}), we can see that the secondary 
bar is significantly weakened by the formation of the primary bar, so that it 
can no longer extend to its co-rotation radius. It takes more than $300$ 
time units for $A_{\rm sec}$ to settle to a steady state, while the change of 
the primary bar amplitude is very small after $t=200$. This S2B structure 
persists to the end of the simulation, lasting for more than $6$ Gyr. 
The two bars are not rigid bodies when they rotate through each other. 
As seen in the insets of Figure \ref{fig:S2Bevo}, the dotted and dashed lines 
match the local maximum or minimum points of bar amplitudes well, and they also 
match the inflection points of $\Delta \phi$ curves in the inset of Figure 
\ref{fig:S2Bphase}. The secondary bar is stronger and rotates 
slower when the two bars are perpendicular, while it is weaker but rotates 
faster when they are parallel (see also DS07). Generally, the secondary bar 
amplitude and pattern speed variations are larger than in the primary bar. 
Such oscillating pattern speeds and amplitudes are consistent with 
{\it loop}-orbit predictions \citep{mac_spa_00}.

\begin{figure}[htp]
        \centering
        \subfigure{\includegraphics[width=0.48\textwidth]{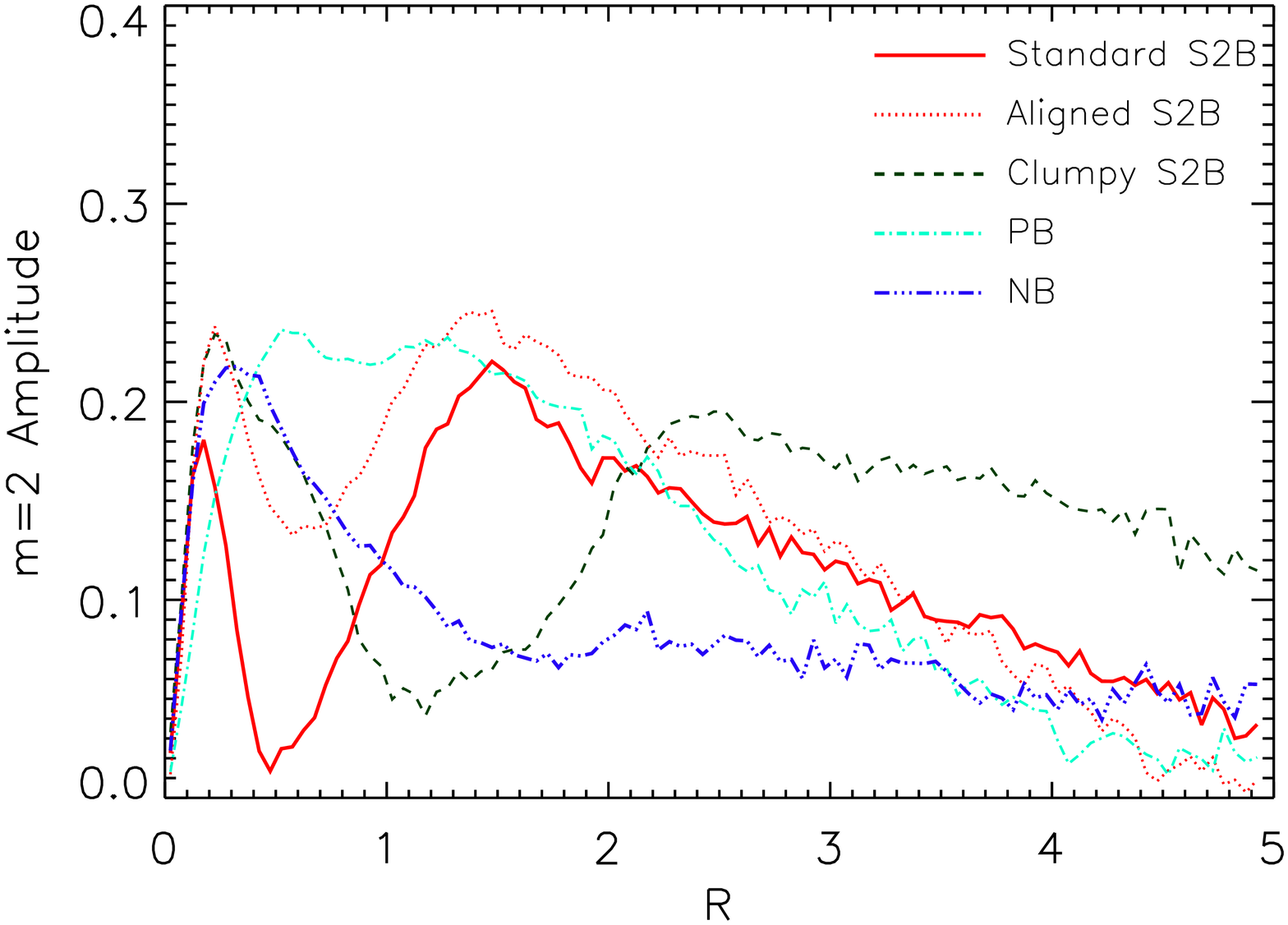}}
        \subfigure{\includegraphics[width=0.48\textwidth]{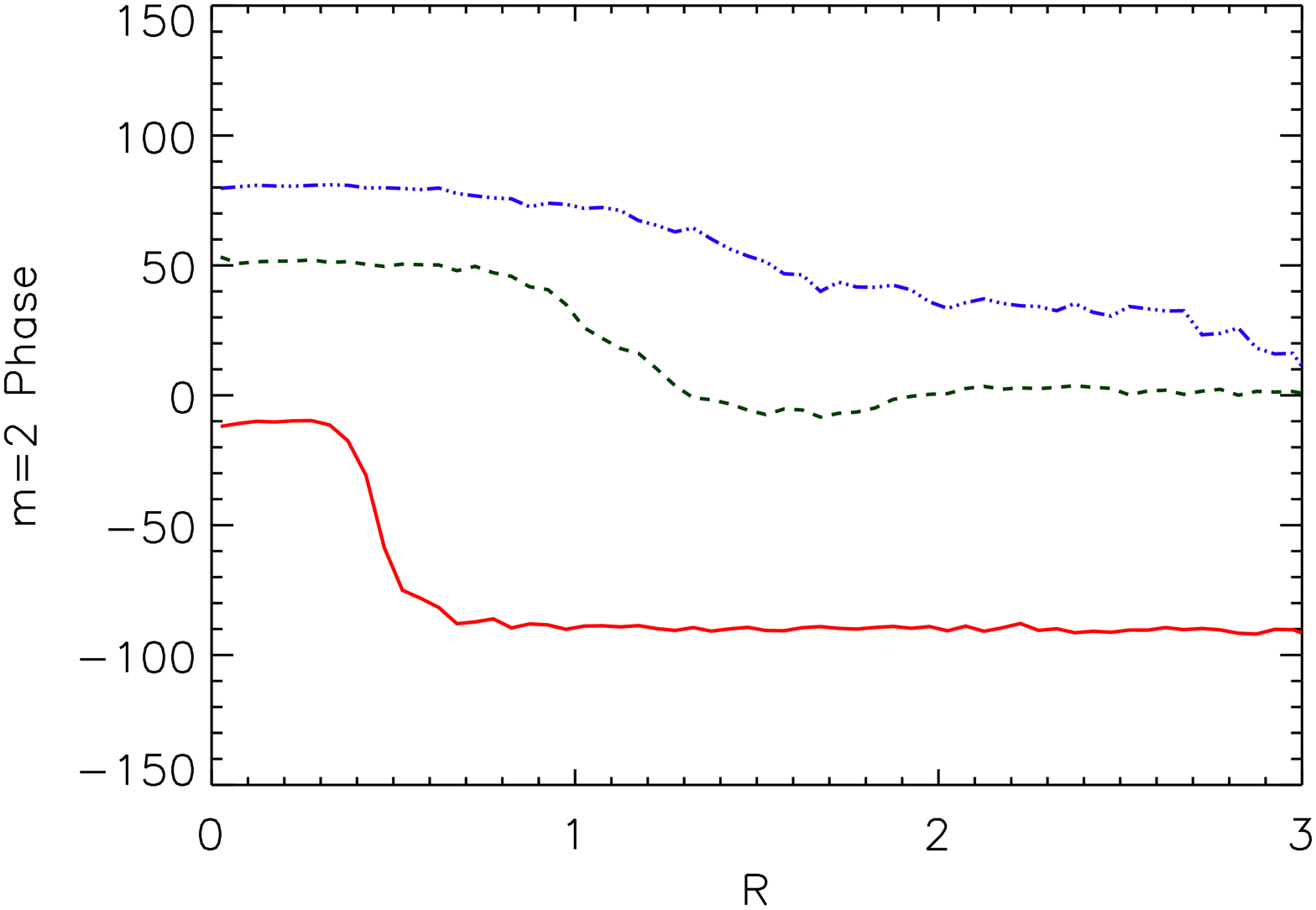}}
    \caption{The $m=2$ amplitude and phase as a function of radius at $t=500$ 
    when all structures are stable. Here we take the simulations discussed in 
    this paper as representative of each type of models to show their bar
    strength and length. The phase of the small-scale bars in the standard S2B, 
    clumpy S2B and NB models are shown in the bottom panel. }
        \label{fig:amplphase}
\end{figure}

In order to compare with observations, we use two methods to measure the 
bar length. As shown in the top panel of Figure \ref{fig:amplphase}, the 
secondary bar peak amplitude of the standard model is at $R \sim 0.2$. The 
length of the secondary bar, $a_{\rm sec}\sim0.3$, is measured by 
tracing half-way down the peak on the $m=2$ amplitude plot, which is 
consistent with the value $0.4$ given by the $10^\circ$ deviation from a 
constant phase. The semi-major axis of the primary bar, 
$a_{\rm prim}\sim3.0$, gives a length ratio  
$a_{\rm sec}/a_{\rm prim}\sim0.10-0.13$ that is consistent 
with the typically observed length ratio of local S2B systems (median ratio 
$\sim 0.12$) \citep{erw_spa_02, erw_04, lis_etal_06}. The primary 
bar extends roughly to its co-rotation radius 
($R_{CR, {\rm prim}}\sim3.0\sim{a_{\rm prim}}$), while the secondary bar 
is much shorter than its co-rotation radius ($R_{CR, {\rm sec}}\sim1.5$). 
If $1.0 \le R_{CR}/a \lesssim 1.4$, the bar is classified as a ``fast'' 
bar \citep[e.g.][]{deb_sel_00, agu_etal_03, cor_etal_03, deb_wil_04}. Therefore, the primary 
bar is a ``fast'' bar, while the secondary bar is ``slow''. 
It is worth noting that the approximate co-rotation radius of the 
secondary bar ($R_{CR, {\rm sec}}\sim1.5$) is close to the outer ILR 
radius of the primary bar ($R_{oILR, {\rm prim}}\sim1.5$), which is 
considered as an evidence of CR-ILR coupling by \citet{rau_sal_99} 
\citep[see also][]{pfe_nor_90, fri_mar_93} 

Moreover, before the formation of the primary bar, the rapidly evolving 
inner disk exhibits recurrent transient spirals driven by the secondary bar 
as seen at $t=20$ in Figure \ref{fig:S2Bform}. As the primary bar grows 
stronger, the nuclear spirals gradually disappear. Therefore, a reasonable 
inference is that the formation of the primary bar efficiently suppresses 
nuclear spirals. 


\subsection{A clumpy double-barred galaxy}
\label{subsectionparspace3}
\begin{figure*}[htp]
        \centering
        \subfigure{\includegraphics[width=0.28\textwidth, angle=-90]{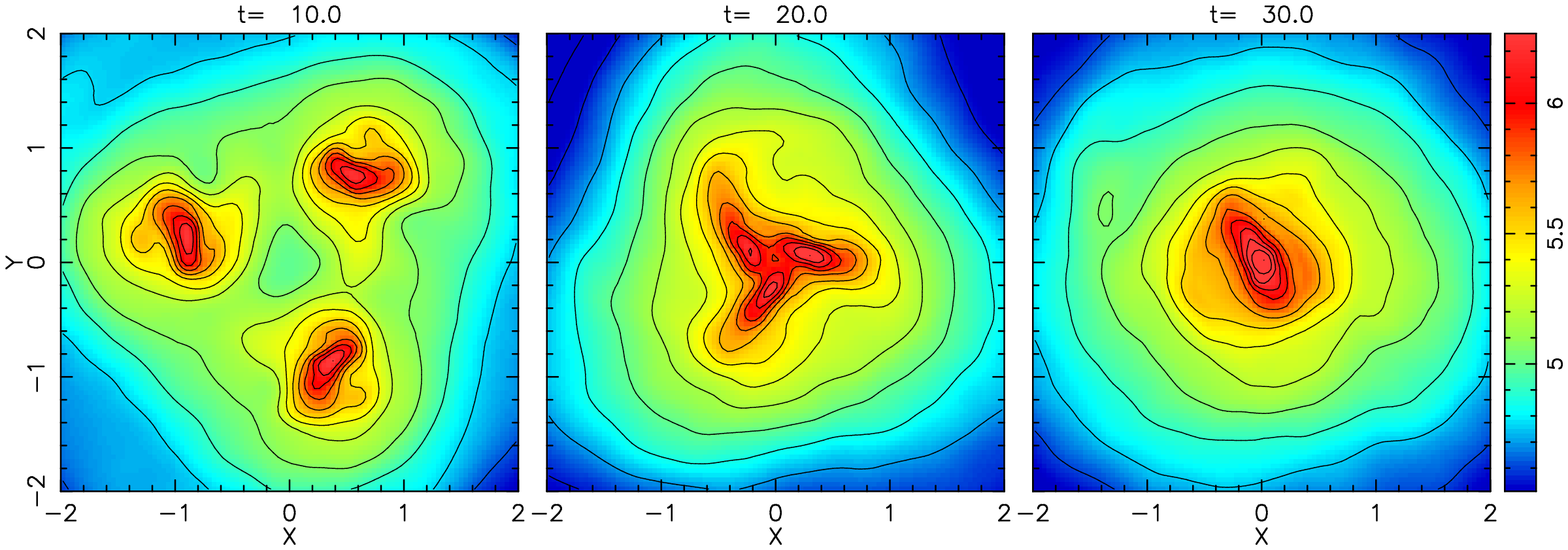}}
        \subfigure{\includegraphics[width=0.28\textwidth, angle=-90]{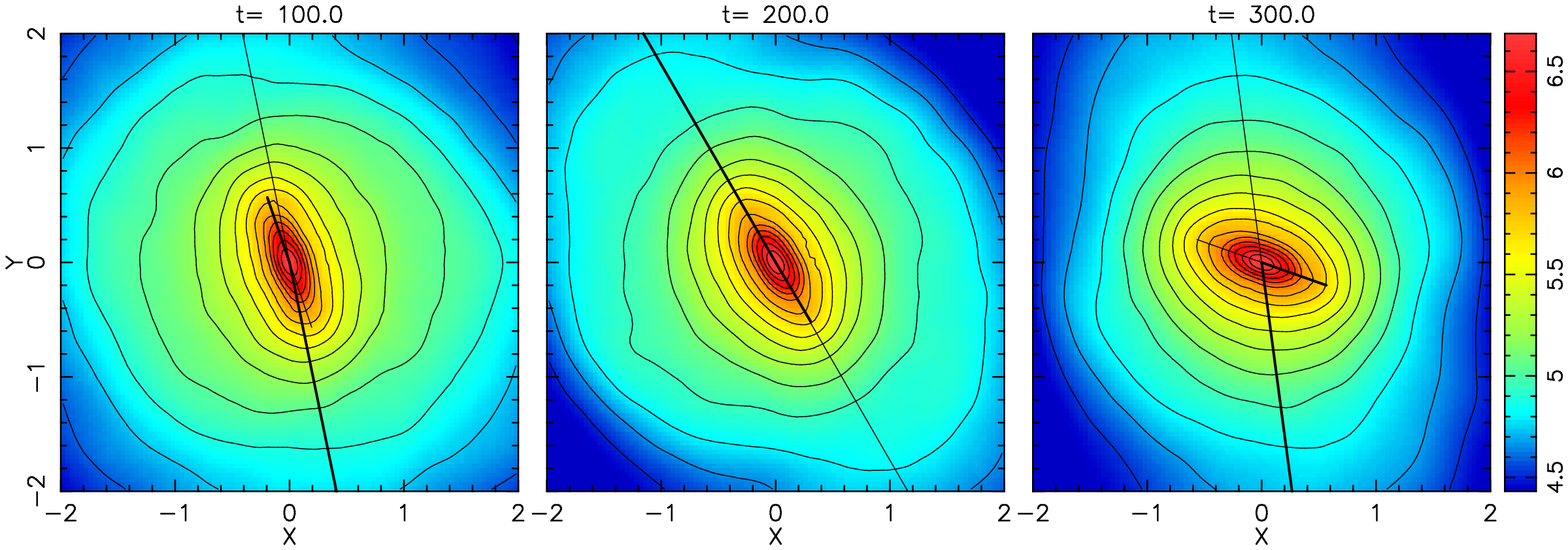}}
        \caption{The formation process of a double-barred galaxy 
        from a clumpy phase. This simulation has $M_d=1.5$ and $b_Q=0.3$.}
        \label{fig:clumpy}
\end{figure*}
\begin{figure*}[htp]
        \centering
        \subfigure{\includegraphics[width=0.2\textwidth, angle=-90]{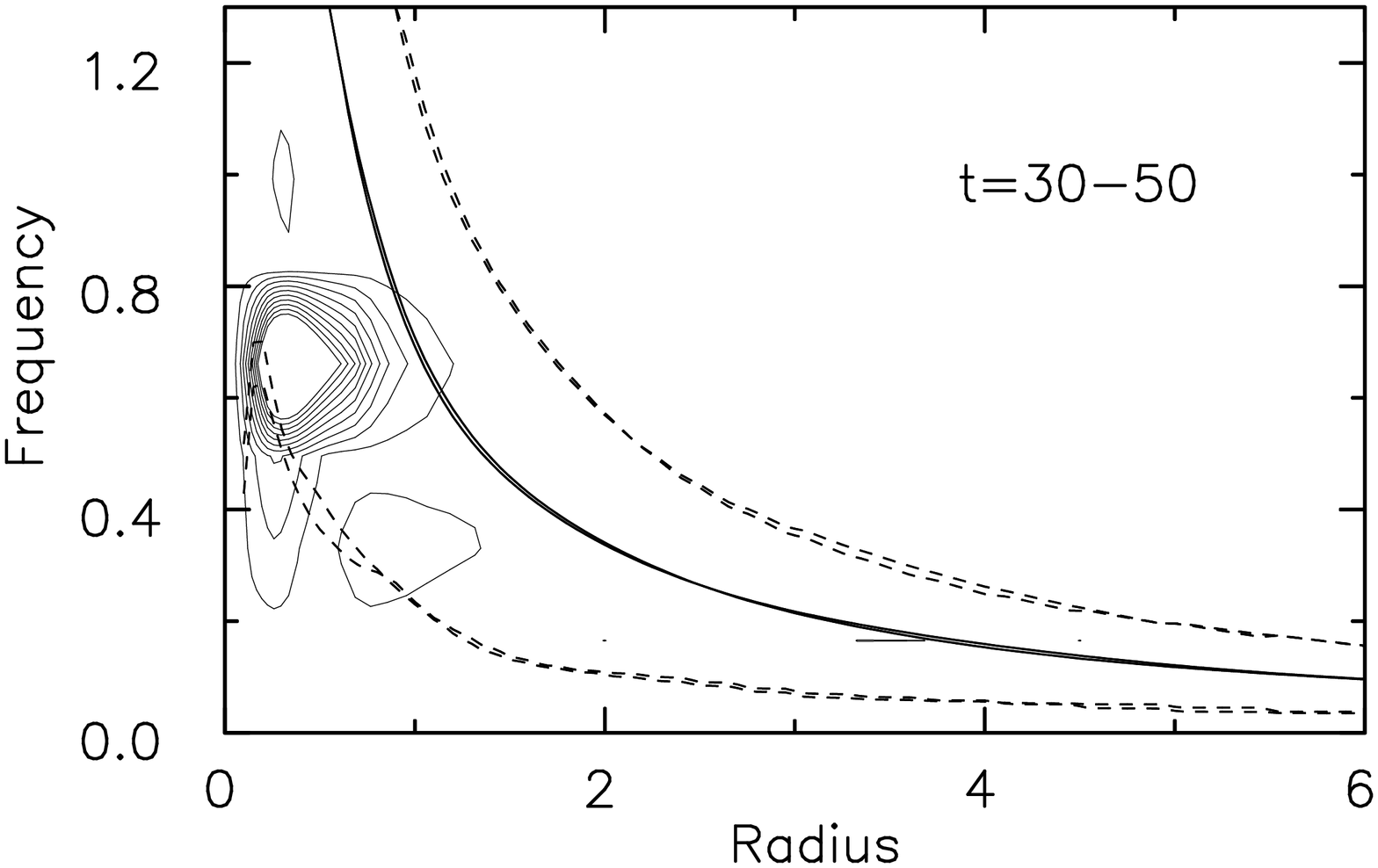}}
        \subfigure{\includegraphics[width=0.2\textwidth, angle=-90]{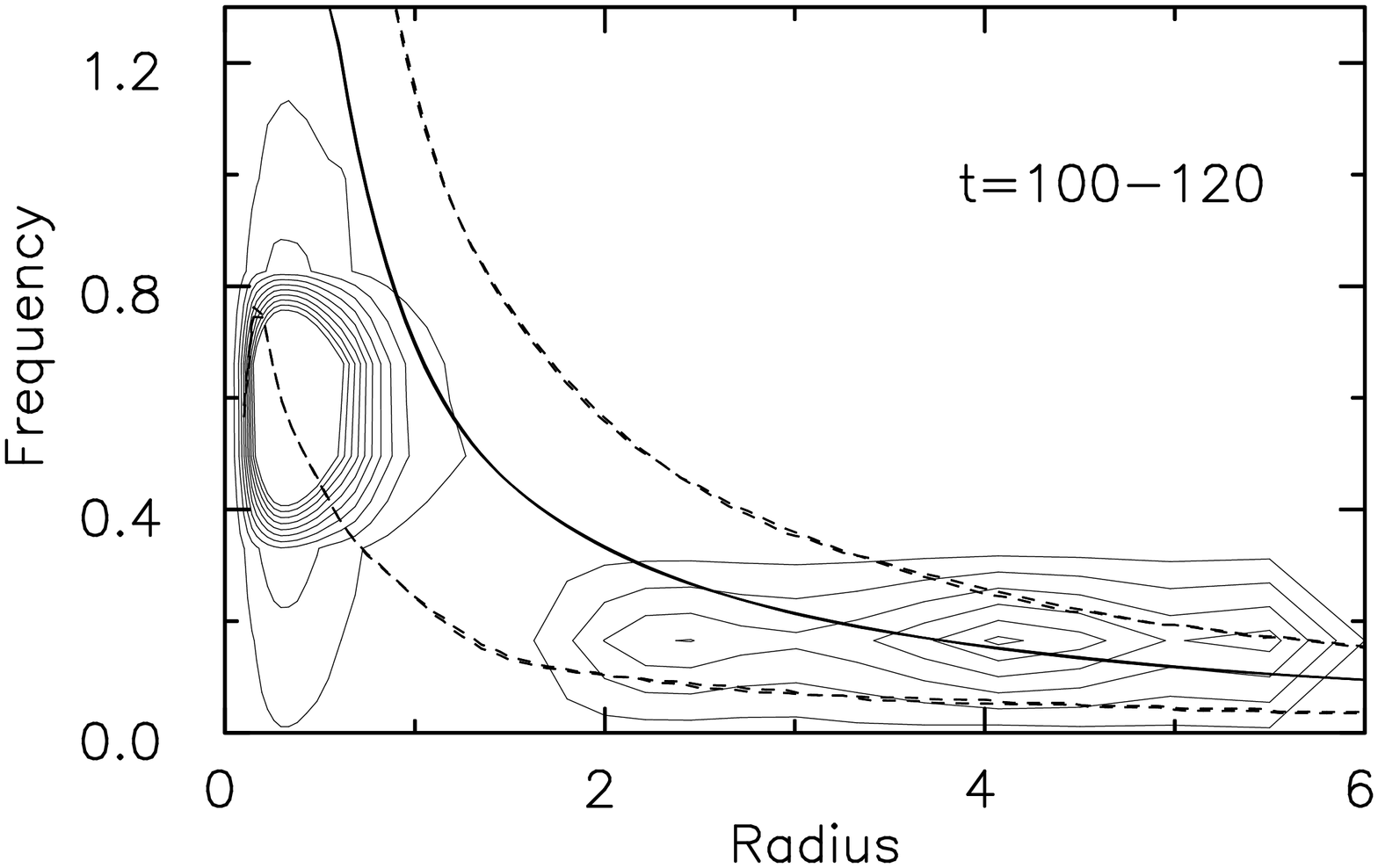}}
        \subfigure{\includegraphics[width=0.2\textwidth, angle=-90]{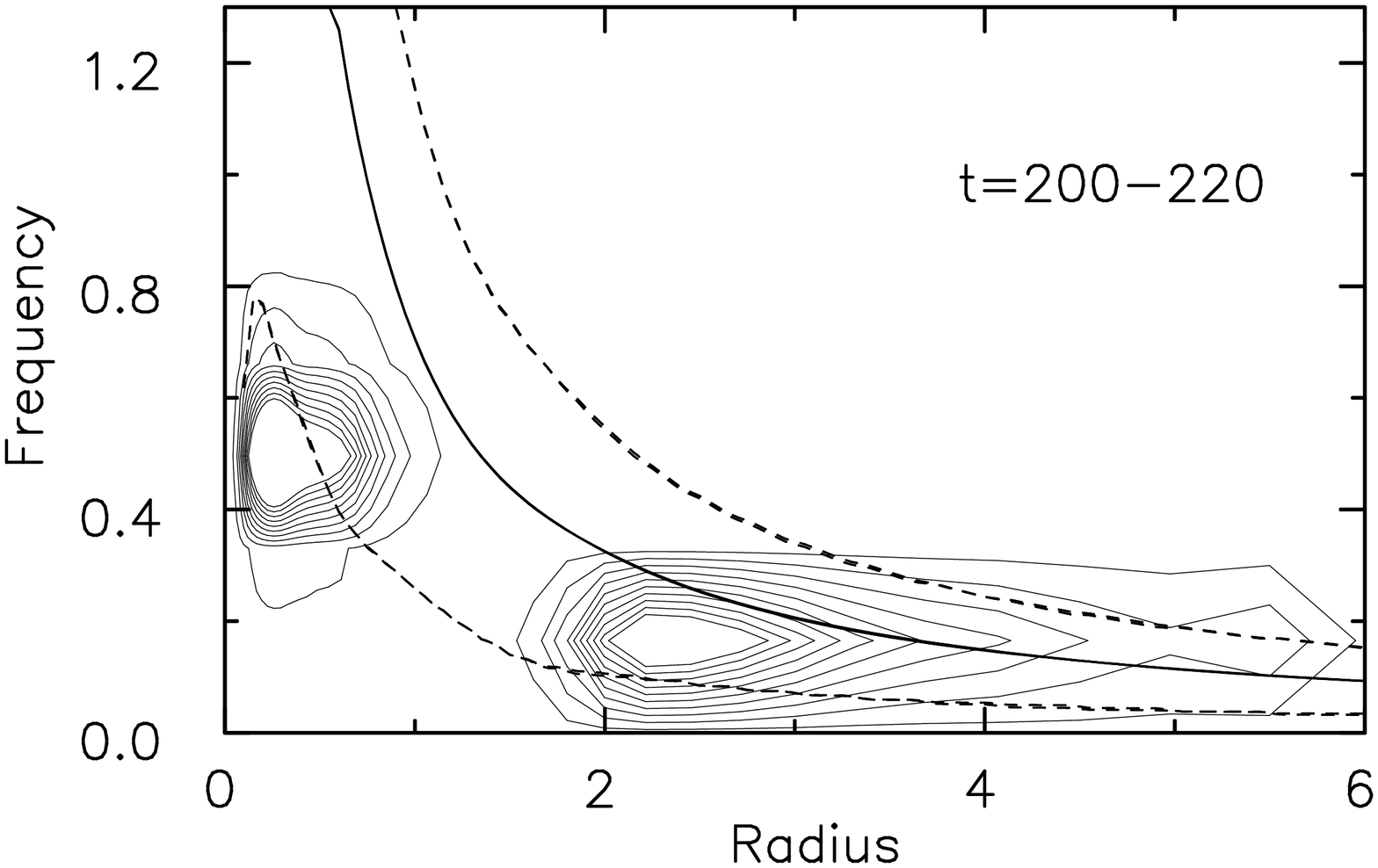}}
        \caption{Power spectra of the $m=2$ Fourier component as a function 
        of radius for the clumpy model in Figure \ref{fig:clumpy}.}
        \label{fig:clumpyfreq}
\end{figure*}
\begin{figure}[htp]
        \centering
        \includegraphics[width=0.45\textwidth]{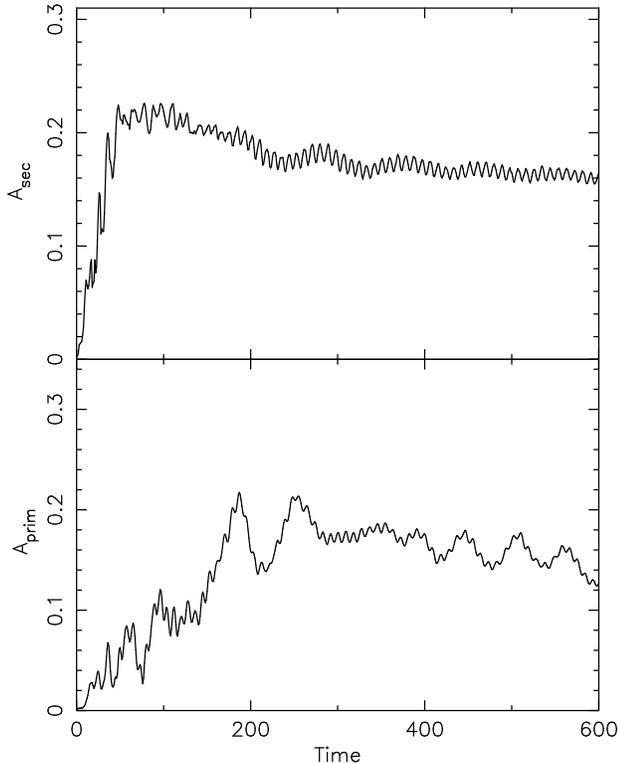}
        \caption{Time evolution of the amplitude of the secondary bar 
        (top panel) and the primary bar (bottom panel) of the clumpy S2B. 
        $A_{\rm prim}$ and $A_{\rm sec}$ are defined as the Fourier $m=2$ 
        amplitude over the radial ranges $2.00\le{R}\le4.07$ and 
        $R\leq0.96$, respectively.}
        \label{fig:clumpyS2Bevo}
\end{figure}

\begin{figure}[htp]
        \centering
        \includegraphics[width=0.48\textwidth]{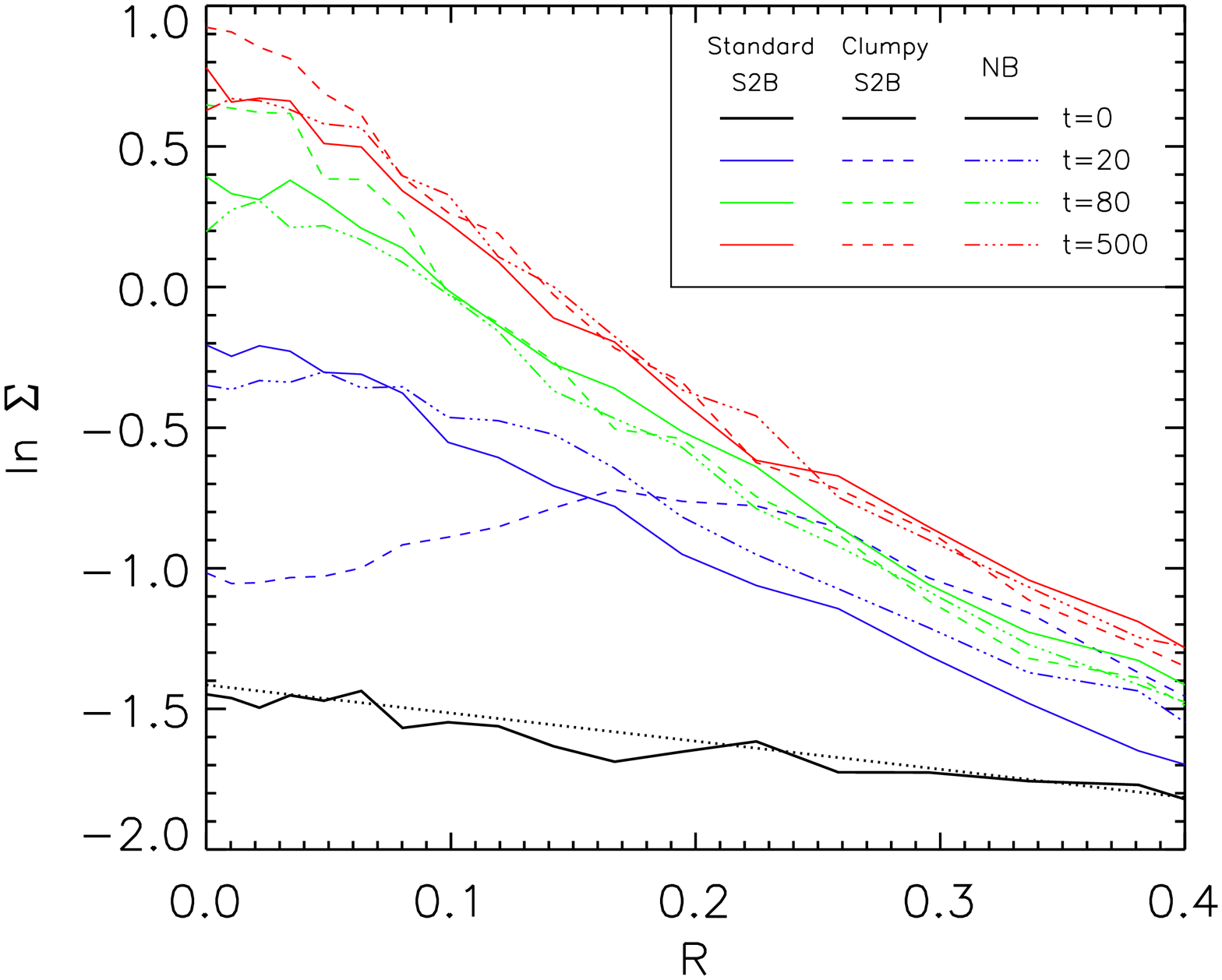}
        \caption{The central surface density profiles of the standard S2B, 
        clumpy S2B and NB models at $M_d=1.5$ highlighted by the dashed 
        black rectangle on Figure \ref{fig:parspace}. The black line is 
        their initial surface density profile.} 
        \label{fig:Surfdens}
\end{figure}

When $b_Q\lesssim0.45$, all simulations experience a violent clumpy phase, 
shown by the shaded regions in Figure \ref{fig:parspace}. As expected, 
with a colder inner disk, the clump instability becomes more significant. 
The clump instability occurs mainly at $R\lesssim0.7$ where the initial 
$Q$ is less than $0.8$, which is roughly consistent with the criterion 
of \citet{dek_etal_09} ($Q\simeq0.67$) for clump growth in an 
unstable disk. A thinner inner disk will also strengthen the clump 
instabilities, but thickness is not as important as $Q$. When 
$b_Q \gtrsim 0.7$, we never see clear clump instabilities even with 
a very thin inner disk. As shown in Figure \ref{fig:clumpy}, the inner disk 
fragments into clumps in a short time. Because of clump-clump gravitational 
interactions and dynamical friction, the massive clumps sink to the center 
and merge to form a small-scale bar. The polar grid code we adopt concentrates 
spatial resolution at the geometric center of the grid. However, the 
formation of clumps results in the highest density not coinciding with the 
region of highest spatial resolution. Therefore, we have tested refining 
the grid by doubling the number of radial and azimuthal grid cells; the 
result of this test is similar to the result shown here. 

As with standard S2Bs, the two bars in clumpy S2Bs also form 
from independent bar instabilities of the inner and outer disks. 
Although the formation of the secondary bar is slightly delayed by the 
violent clumpy phase, as shown in Figure \ref{fig:clumpyfreq}, the newly 
formed secondary bar can extend to its co-rotation radius as well, 
$R_{CR, {\rm sec}} \sim 1.0$. In our clumpy simulations, the clumpy phase 
always results in the formation of a small-scale bar at the beginning.  
However, they differ from standard S2Bs in terms of their dynamical 
properties. Firstly, the amplitude evolution of the primary bar (Figure 
\ref{fig:clumpyS2Bevo}) shows that it forms much later and more gradually
in the clumpy S2B. Secondly, the iso-density contours (Figure \ref{fig:clumpy}) 
show that the secondary bar is more disky and central concentrated, 
while the secondary bars of standard S2Bs tend to be more rectangular-shaped. 
We perform ellipse fitting for both models using \texttt{IRAF}, and find 
more negative $B_4$ for the secondary bar of the standard S2B. In Figure 
\ref{fig:Surfdens}, we show the central surface density profiles of the three 
models. Because they have the same disk mass $M_d=1.5$, they share the same 
density distribution at $t=0$. After the small-scale bar forms, at $t=80$, the 
central 
surface density of the clumpy S2B is $\sim35\%$ larger than both the standard 
S2B and NB model. From $t=80$ to $t=500$, the surface density of all these 
three models increases at the same rate. Therefore, the extra central 
surface density of the clumpy S2B is produced during the clumpy phase. 
Thirdly, the two bars of the clumpy S2B are both much longer than those of 
the standard S2B as shown in Figure \ref{fig:amplphase}. Measured by 
tracing half-way down the slope of $m=2$ amplitude, the length ratio is 
$a_{\rm sec}/a_{\rm prim}\sim0.8/5.0=0.16$ that is slightly larger than 
standard S2Bs. If the bar length is measured at the radius where the $m = 2$ 
phase deviates from a constant by $10^\circ$, $a_{\rm sec}$ is even larger 
at $1.0$. Measured from Figure \ref{fig:clumpyfreq} ($t=200-220$), 
$R_{CR, {\rm sec}}$ is about $1.3$, then 
$R_{CR, {\rm sec}}/a_{\rm sec} \sim 1.3-1.6$. 
Thus the secondary bar is a quite ``fast'' bar extending close to its 
co-rotation radius, which is quite common in our clumpy S2Bs. This 
surprising result demonstrates that it is dynamically possible to 
generate a stable ``fast'' double-barred system. 
As shown in the top panel of Figure \ref{fig:clumpyS2Bevo}, the change of 
$A_{\rm sec}$ is quite small with the formation of the primary bar. Moreover, 
the two bars oscillate in the same way as standard S2Bs, but the oscillation 
is smaller. These indicate the interaction of two bars is weaker. In Figure 
\ref{fig:clumpy} ($t=200, 300$), we can clearly see that the inner disk is 
dominated by the secondary bar. Thus the primary bar is not efficient at 
trapping the secondary bar in the clumpy S2B models. A possible 
reason is that the torques from the primary bar are not strong enough to 
reduce the strength and length of the secondary bar, because the primary 
bar is much fatter and less massive than the ones of standard S2Bs. 
This is because the primary bar forms further out in the disk. Moreover, a 
more massive and concentrated secondary bar may also stabilize it against 
the primary bar. As shown in Figure \ref{fig:Surfdens}, in the very central 
region the clumpy S2B is more massive than the standard S2B. In conclusion, 
in our clumpy S2Bs, the primary bar is too weak to be efficient at trapping 
the secondary bar, in which case a ``fast'' double-barred system can be 
sustained for a long time.

These clumpy S2B simulations demonstrate that a small-scale bar can be generated 
from a violent clumpy phase. Previous N-body+gas simulations showed that 
clumps coalesce into a bulge whose shape was bar-like 
\citep[][Figure 2 and 3]{elm_etal_08}. \citet{ino_sai_12} showed that  
clumpy-origin bulges are bar-like in their hydrodynamic (SPH) simulations, 
and have exponential surface density profile, boxy (bar-like) shape 
and significant rotation. All these properties are consistent with the 
small-scale bars in our simulations. Therefore, these ``bar-like'' 
structures may be similar to our small-scale bars, which suggests that 
the secondary bar might often form from clump mergers. 

\subsection{An aligned double-barred galaxy}
\label{subsectionparspace2}
\begin{figure*}[htp]
        \centering
        \subfigure{\includegraphics[width=0.28\textwidth, angle=-90]{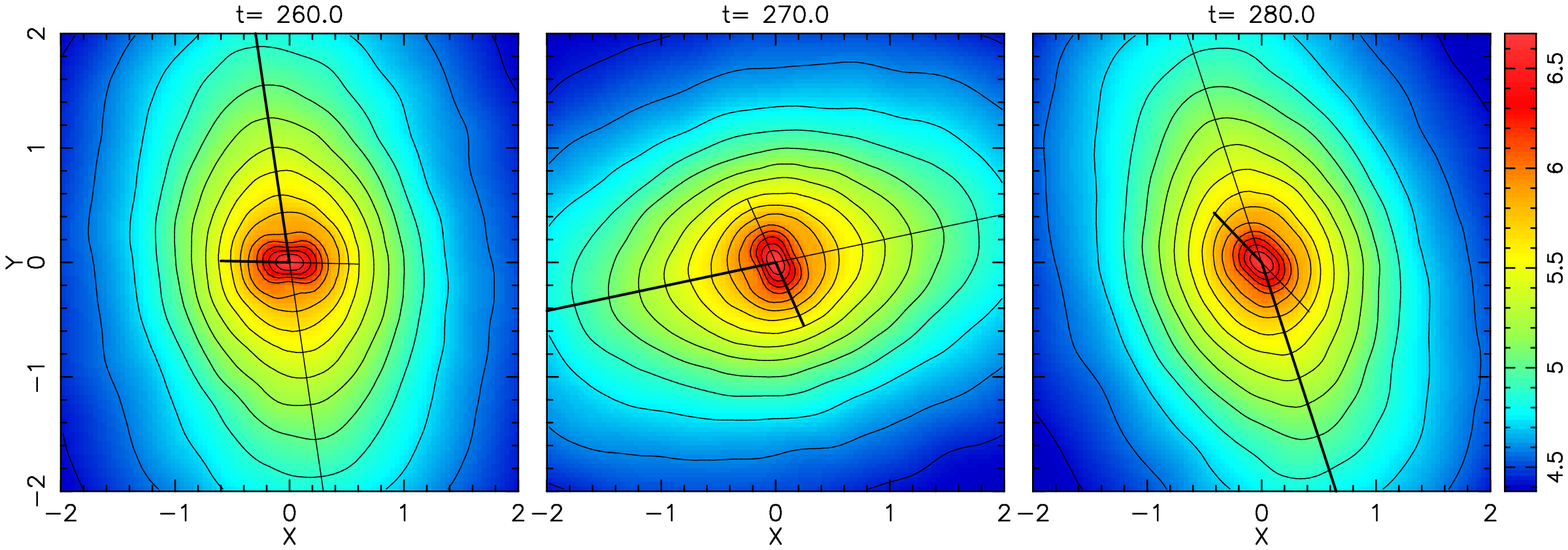}}
        \subfigure{\includegraphics[width=0.28\textwidth, angle=-90]{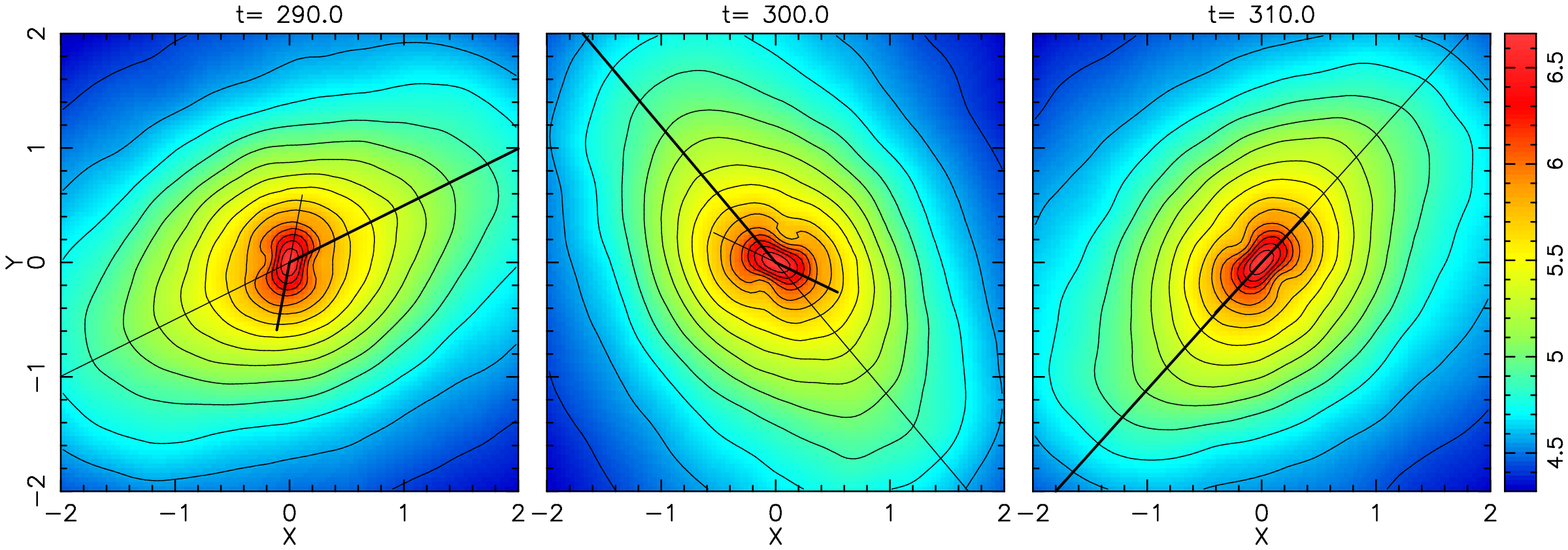}}

        \caption{The coupling phase of the two bars in the aligned S2B case. This 
        simulation has $b_Q=0.5$ and $M_d=1.1$. Note the developing of peanut-shaped 
        contours from $t = 290$. A movie showing the coupling behavior in the corotating frame of the primary 
        bar is available at \href{http://hubble.shao.ac.cn/~dumin/S2B/Video2.gif}{hubble.shao.ac.cn/$\sim$dumin/S2B/Video2.gif} and the ApJ website}
        \label{fig:couple}
         \subfigure{\includegraphics[width=0.2\textwidth, angle=-90]{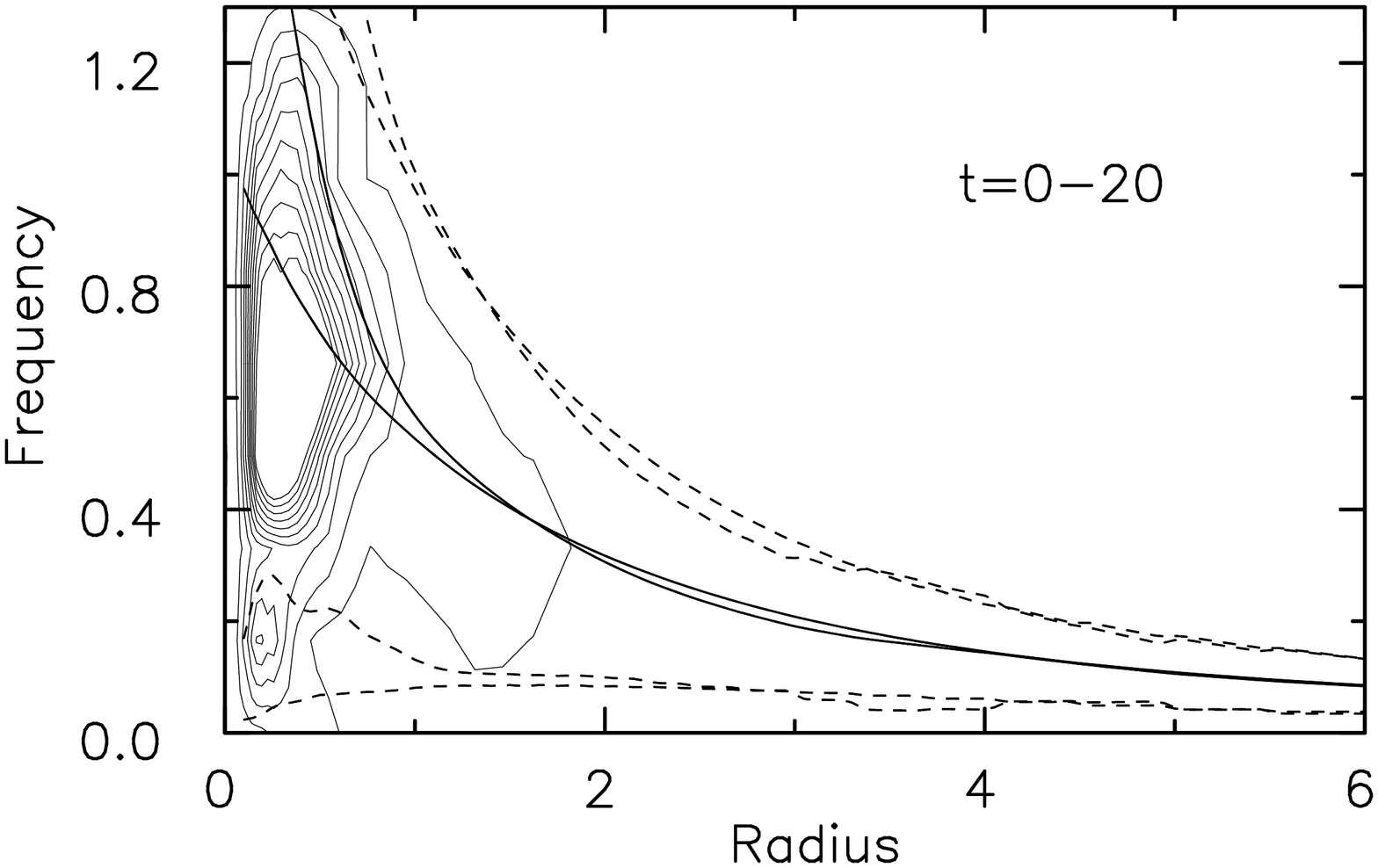}}
         \subfigure{\includegraphics[width=0.2\textwidth, angle=-90]{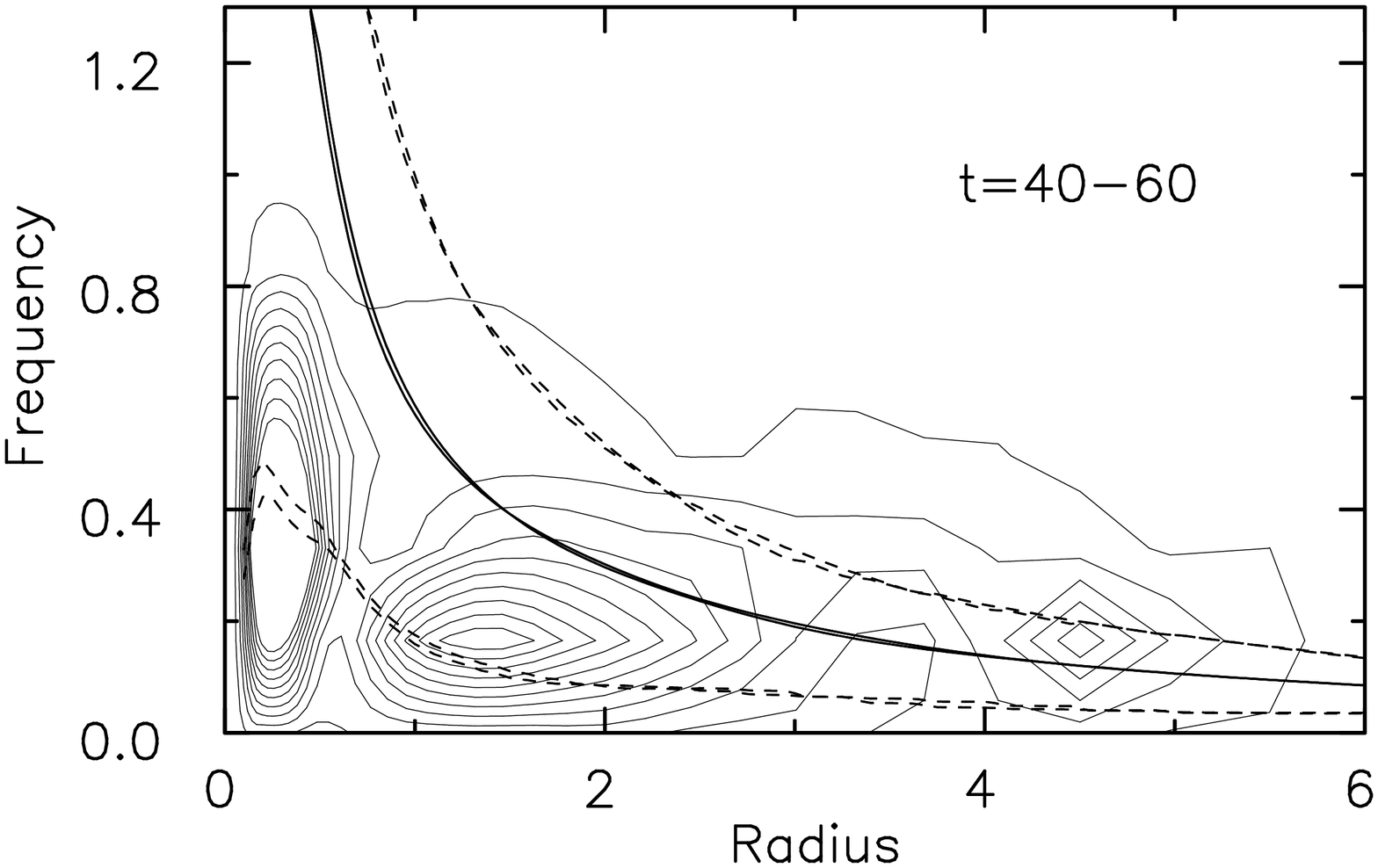}}
         \subfigure{\includegraphics[width=0.2\textwidth, angle=-90]{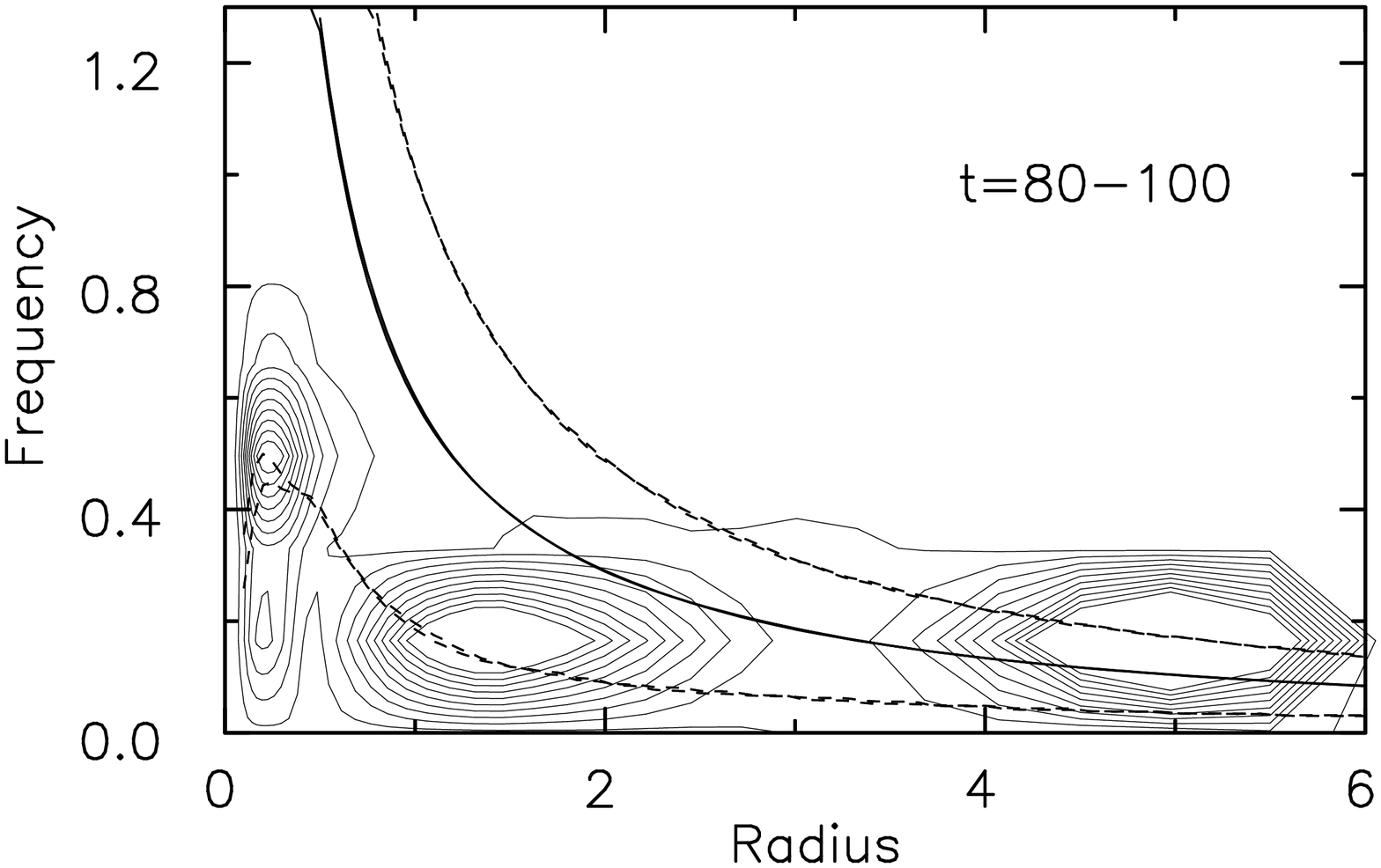}}
         \subfigure{\includegraphics[width=0.2\textwidth, angle=-90]{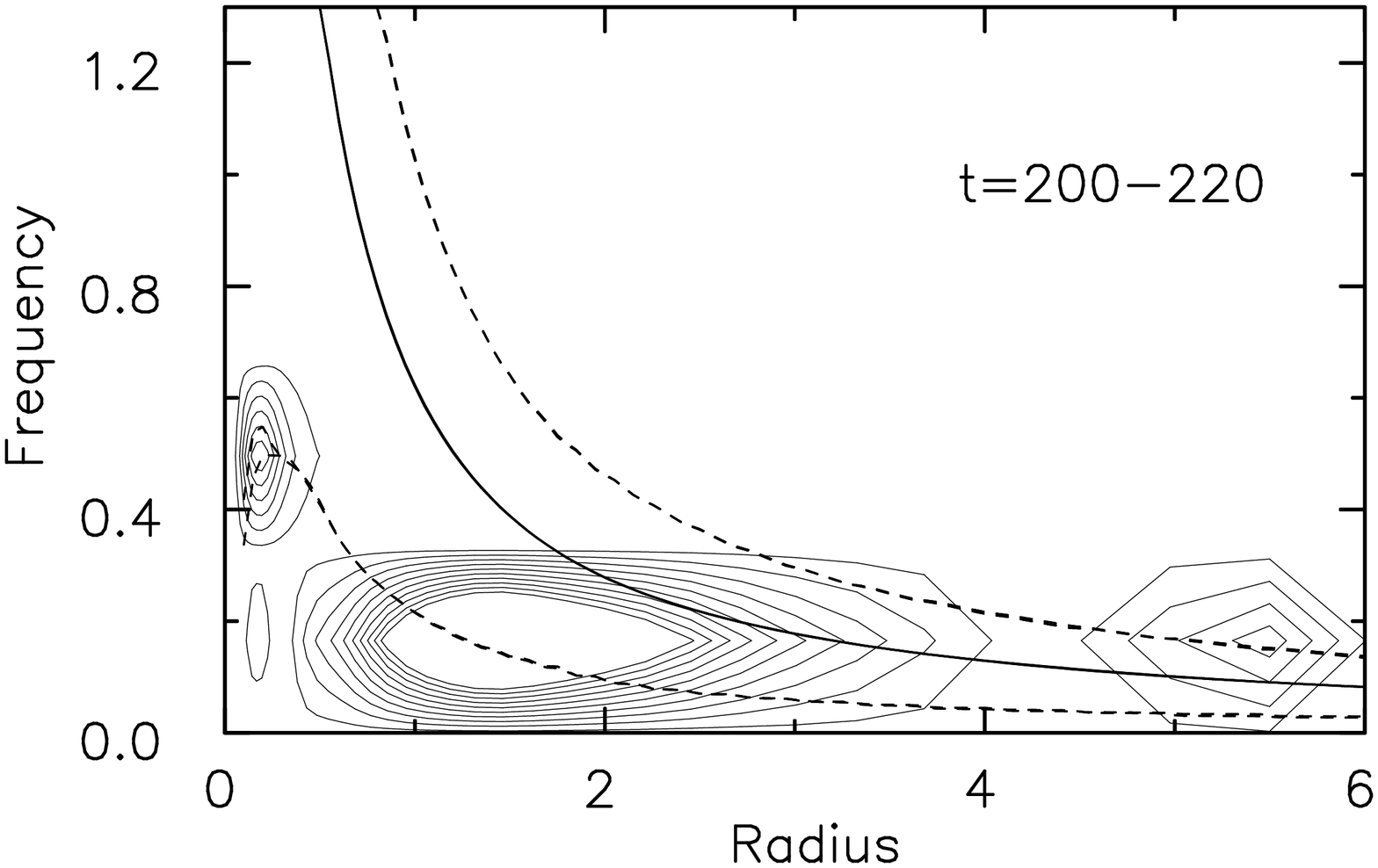}}
         \subfigure{\includegraphics[width=0.2\textwidth, angle=-90]{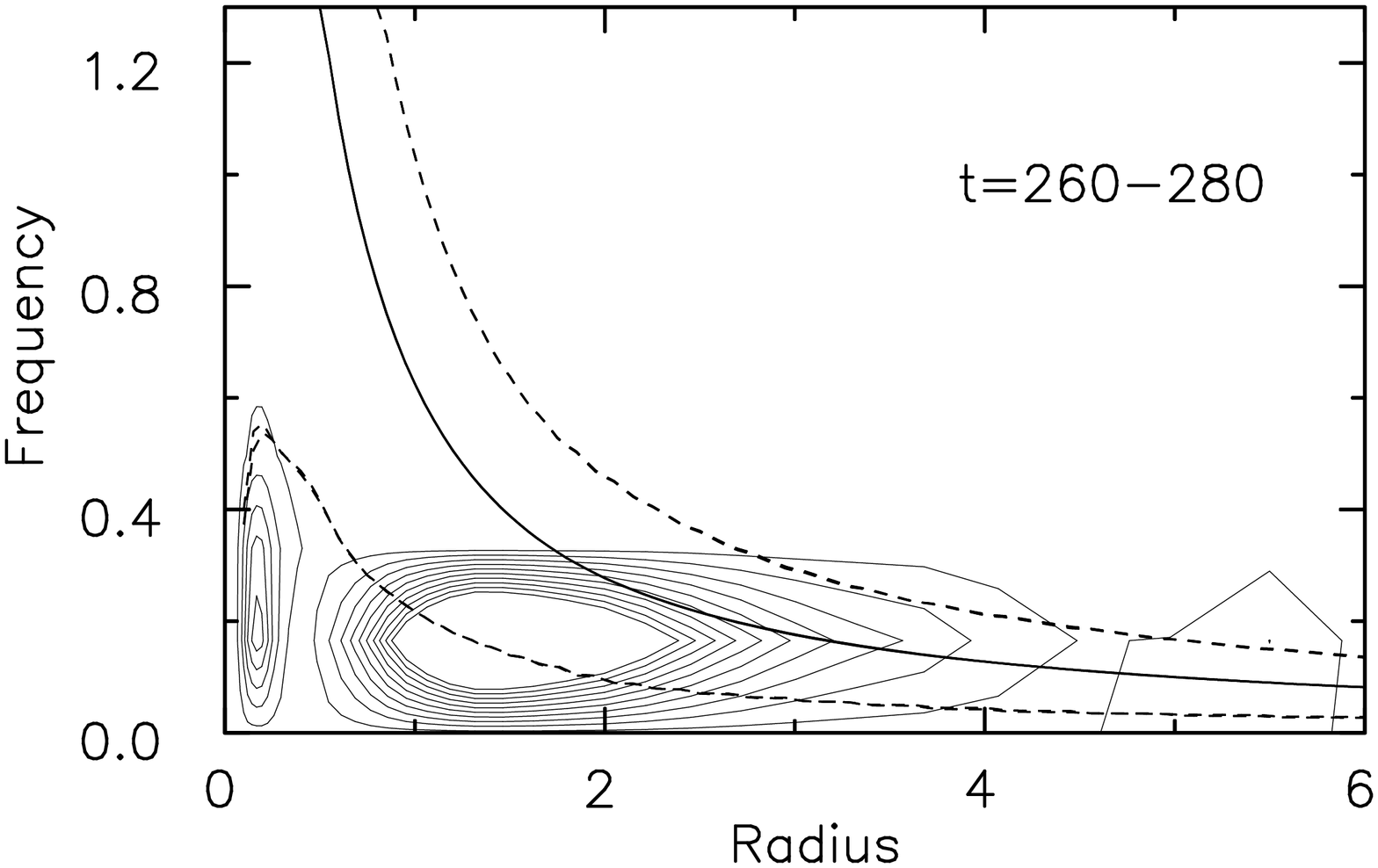}}
         \subfigure{\includegraphics[width=0.2\textwidth, angle=-90]{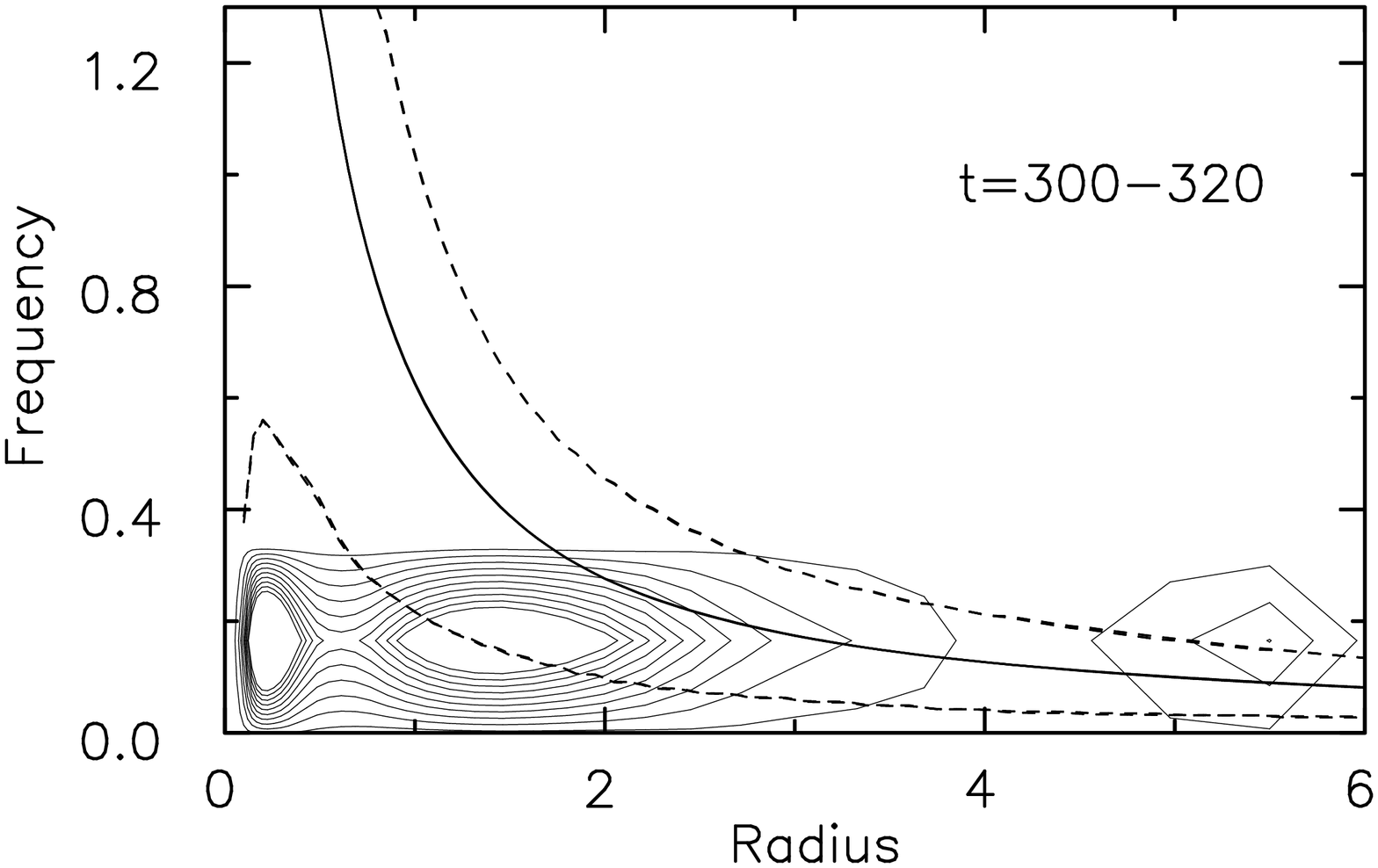}}
         \caption{Power spectra of the $m=2$ Fourier component as a function of 
             radius for the aligned S2B model in Figure \ref{fig:couple}.}
         \label{fig:couplefreq}
\end{figure*}
\begin{figure}[htp]
        \centering
        \includegraphics[width=0.48\textwidth]{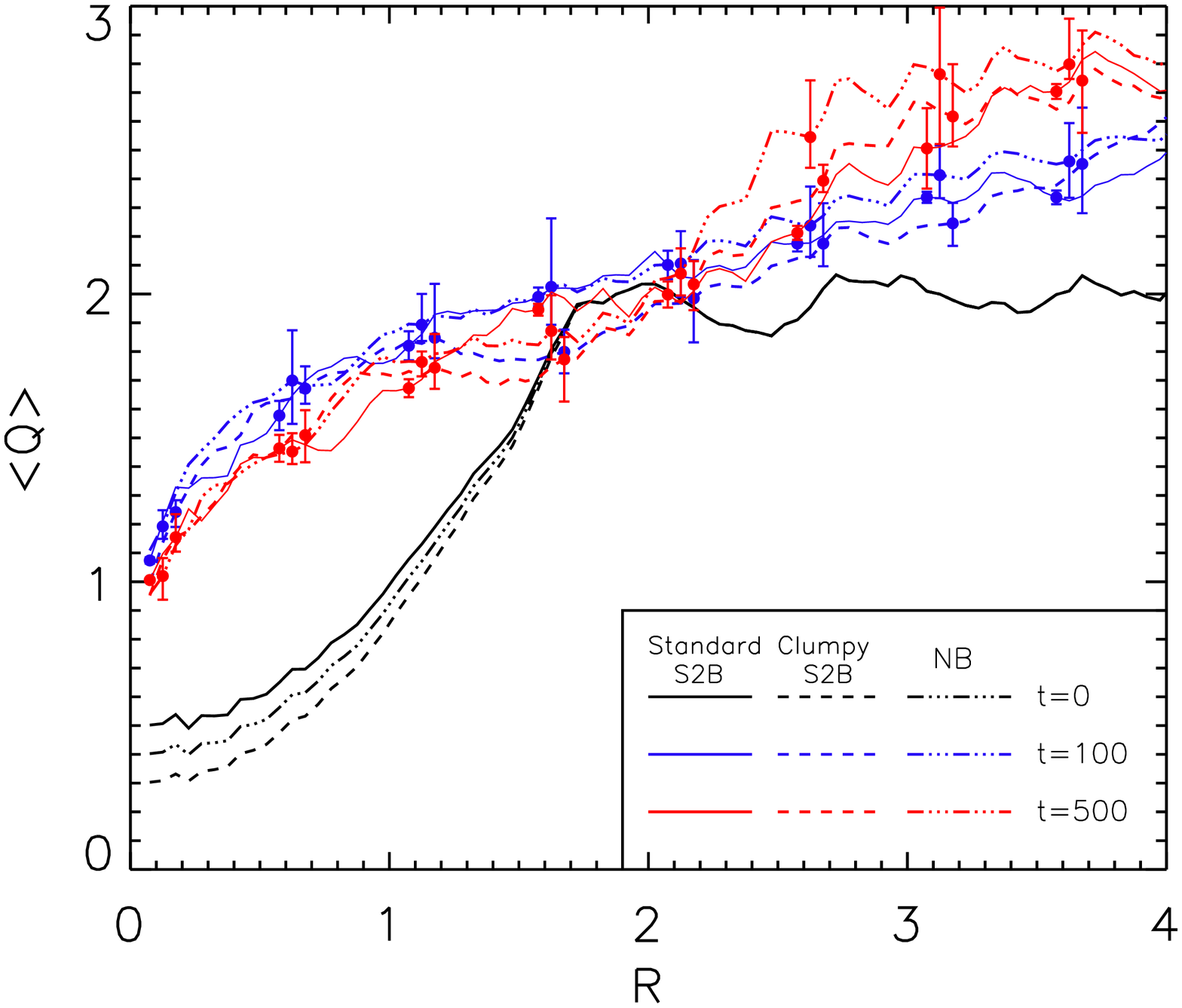}
        \caption{The evolution of the mean value of $Q$ of standard S2Bs, 
            clumpy S2Bs, and NBs that correspond to the three 
            groups of models in Figure \ref{fig:parspace} within 
            the red, blue and green rectangles, respectively. The black
            lines are the initial $Q$ profiles of these simulations. 
            The error bars refer to the scatter of each group of 
            simulations.}
        \label{fig:ToomreQevo}
\end{figure}

Figures \ref{fig:couple} and \ref{fig:couplefreq} show the formation and 
coupling process of an aligned S2B. As shown in Figure 
\ref{fig:couplefreq}, the initial bar instabilities form two bars with 
different pattern speeds between $t=0$ and $80$. The oscillating properties 
are identical to those of the standard and clumpy S2Bs. During the period 
$t=80-270$, the secondary bar gradually becomes shorter and weaker because 
of the interaction with the primary bar. At $t \simeq 280$ (Figure 
\ref{fig:couple}) the secondary bar cannot pass through the next 
perpendicular barrier where the rotational kinetic energy of the secondary bar 
is minimal due to the gravitation torque from the primary bar. Then the 
pattern speed of the secondary bar decreases sharply to that 
of the primary bar. The secondary bar remains at fixed orientation relative 
to the primary bar during $t=280-300$. The two bars share the same pattern 
speed, although they are still mis-aligned at a roughly constant angle. 
During this time, the surface density contours around the secondary bar 
become somewhat peanut-shaped. This shape transformation always happens 
in all our aligned S2Bs. Then the secondary bar falls back to alignment with 
the primary bar gradually, which is the lowest energy state. Sometimes, the 
secondary bar librates slightly around the major axis of the primary bar 
before complete alignment. Finally, they rotate at the same pattern speed 
until the end of the simulation. Generally, this coupling period occurs 
over a few tens of time units ($\sim 100$ Myr). 

Although the aligned S2B now resembles a single-barred galaxy, the 
relic of the secondary bar can be clearly seen as a nuclear peak in the 
$m=2$ amplitude profile (Figure \ref{fig:amplphase}). The peanut-shaped 
contours are another signature of aligned S2Bs that survive to the end 
of our simulations. Normally, the coupling process 
is dominated by the primary bar as the secondary bar gradually becomes 
shorter and weaker. Only very few models show a coupling process dominated 
by the secondary bar, in which the primary bar is nearly destroyed by the 
secondary bar during their coupling process. A fraction of the particles 
previously belonging to the primary bar are, in this case, trapped around 
the secondary bar, making it longer. 

For our standard S2Bs, the pattern speed ratio of the two bars, 
$\Omega_{\rm prim}/\Omega_{\rm sec}$, varies from $0.36$ to $0.55$, while 
it is $0.25-0.43$ for the clumpy S2Bs. Generally, the pattern speeds of the 
primary bars of the clumpy S2Bs are lower than of the standard S2Bs. Before 
alignment, the pattern speed of the aligned S2Bs is $0.35-0.49$, i.e. 
similar to the range of the standard S2Bs. Therefore, it is hard to predict 
whether a S2B system will couple or not based only on its pattern speed 
ratio. Possibly, the mass and morphology of two bars also play a role.  

\section{Large-scale single-barred galaxies and nuclear-barred galaxies}
\label{sectionsingle}

When $b_Q \geq 0.8$, most of the simulations form large-scale single-barred 
galaxies, PBs, as can be seen in Figure \ref{fig:parspace}. Compared with the 
formation process of S2Bs, the bar instability starts with a lower 
pattern speed. 
By losing angular momentum to the outer disk, the bar gradually slows down. 
Then its co-rotation radius is pushed further out and more particles at
larger radius can be trapped forming a longer bar \citep{ath_03}. 
In this way, they cannot form two independently rotating bars.

A fraction of simulations only leave small-scale bars after their weak primary 
bars are gradually dissolved, or, in some cases, never formed at all. These 
models are marked as ``NB'' 
(nuclear-barred model) in Figure \ref{fig:parspace}. Although NBs are also a 
type of single-barred galaxy, compared with the PBs, they have different 
formation histories and properties. To form NBs, a mechanism is needed to 
prevent the mass of the outer disk from being trapped into the primary bar. 
The initial $Q$ of the outer disk is about $2.0$, which is close to the 
criterion ($2.0-2.5$) for non-axisymmetric stability in all disk 
mass distributions \citep{ath_sel_86}. To quantify how sensitive the 
formation of the primary bar is to the change of $Q$, we fix the $Q$ profile 
of the inner disk, and vary $Q$ in the outer disk from $2.0$ to $2.3$. If 
$Q\geq2.2$, the outer disk is unresponsive to bar instabilities, showing that 
bar formation is very sensitive to $Q$ when $Q$ is large. Even a small 
enhancement of $Q$ in the hot outer disk can significantly suppress the 
formation of the primary bar. 

For the subsamples of NBs, standard S2Bs and clumpy S2Bs, we show the 
evolution of their azimuthally averaged $Q$, $\langle Q \rangle$, in Figure 
\ref{fig:ToomreQevo}. 
From $t=100$ to $t=500$, the changes in the inner disks are very small 
while the outer disks of the NBs become hotter than both standard and clumpy 
S2Bs. Such dynamically hot outer disks become unsuitable to form a steady
primary bar. In some cases, a weak and sometimes spiral-like primary bar forms 
in the early phase, but it is hard to stabilize for a long time. One possible 
explanation for the enhancement of $Q$ is the spirals driven by the nuclear bar.
A lot of studies have shown that spirals efficiently heat disks to more eccentric 
orbits \citep{ber_lin_96, bar_wol_67, car_sel_85, bin_lac_88, jen_bin_90, de_sim_04, min_qui_06}.
In our simulations, we do see stronger spirals recur more frequently in NBs 
than in standard and clumpy S2Bs. 

\section{Discussion and conclusions}
\label{sectionconclusion}

We have been successful in forming double-barred galaxies with a set 
of simple initial conditions involving a cooler inner disk. By setting 
up an increasingly cooler disk towards the center, a S2B structure can 
form naturally. Recent N-body+gas simulations by \citet{woz_15} 
showed that the star formation in the central regions was responsible for 
stabilizing the nuclear bars. New stars formed a dynamically cool 
inner disk where nuclear bar instabilities arose, consistent with our 
results. \citet{rau_sal_99} and \citet{rau_etal_02} found that the secondary 
bar formed from the highly rotating central mode, and models with steeply 
rising rotation curves tended to have a long-lasting secondary bar. 
DS07 used a pre-existing rapidly rotating pseudo-bulge to generate S2Bs. 
All these simulations suggest that a rapidly rotating nuclear component might 
plausibly generate a small-scale bar with high pattern speed. The two bars 
form from independent bar instabilities of the inner and outer parts 
of disks. Thus the pattern speeds of the two bars differ significantly in 
such S2Bs. \citet{sah_mac_13} generated a S2B in a galaxy model whose 
gravitational potential was dominated by the dark matter halo. The two bars 
rotated at comparable pattern speed, but they developed from two independent 
bar instabilities as well. Therefore, we conclude that S2Bs can form by simply 
generating independent bar instabilities in the inner and outer disks, although 
the survival of S2B systems is stochastic because of the chaotic nature of 
disks. A dynamically cool inner disk embedded in a hotter outer disk is an 
easy and natural way to satisfy this condition.

In the early phase of the formation of S2Bs, the secondary bars extend to 
their co-rotation radius, which indicates that they form from the usual bar 
instabilities as with the primary bars. The angular momentum exchange among 
the hosting disk, 
secondary bar and primary bar determines the strength, morphology and final 
pattern speeds of the two bars. The primary bars extend roughly to their 
co-rotation radius, so they are ``fast'' bars. For the standard S2Bs, 
although the secondary bars extend to the co-rotation radius at formation, 
they shrink to a much smaller size to become ``slow'' 
bars because of the suppression from the later-formed primary bars. The 
clumpy S2Bs show that the secondary bar can form from the violent clumpy 
phase. With a weaker primary bar, the clumpy S2B can host a ``fast'' secondary 
bar. Because of the interaction of the two bars and the disk, there are 
also a large fraction of unstable S2Bs, with the two bars coupling to alignment 
in just a few Gyr. 

In our models, the secondary bars always form earlier than the primary 
ones, which is consistent with previous collisionless simulations 
\citep[e.g.][DS07]{rau_etal_02}. However, because the whole galaxy is 
set up at the beginning, we cannot constrain the cold material accumulation of 
the cool inner disk. It is not entirely clear to us how a dynamically cool 
inner disk, required in forming S2Bs, may arise. It is generally thought that 
gas is driven from outside in (e.g. a possible channel is provided by the primary 
bar). In this case gas accumulated near the center can give rise to a cool disk 
that may be able to form a nuclear bar. However, an alternative formation 
scenario is that the secondary bar forms from the violent clumpy phase in the 
early universe when the primary bar is still not formed. In many N-body+SPH 
simulations, these clumpy-origin small-scale bars could be easily mistaken 
for bulges. Therefore, the question remains whether the small-scale bar 
instabilities happens at the early time of the galaxy formation or after 
the formation of the primary bar. To better understand the formation of S2Bs, 
further numerical simulations are required.

\begin{acknowledgments}
We thank Bruce Elmegreen and Witold Maciejewski for insigtful comments and 
discussions on the manuscript.
Min Du thanks Jeremiah Horrocks Institute of the University of Central 
Lancashire for their hospitality during a three month visit while this paper 
was in progress. Hospitality at APCTP during the 7th Korean Astrophysics 
Workshop is kindly acknowledged. 
The research presented here is partially supported by the 973 Program of China 
under grant no. 2014CB845700, by the National Natural Science Foundation of 
China under grant nos.11333003, 11322326, 11073037, and by the Strategic Priority 
Research Program ``The Emergence of Cosmological Structures'' (no. XDB09000000) 
of Chinese Academy of Sciences. This work made use of the super-computing 
facilities at Shanghai Astronomical Observatory.
V.P.D. is supported by STFC Consolidated grant \# ST/J001341/1.
\end{acknowledgments}

\end{document}